\documentclass[12pt, a4paper]{article}
\usepackage[height=23.5cm,width=16.5cm,centering]{geometry}
\usepackage[dvipdfmx]{graphicx}
\usepackage[usenames]{color}
\usepackage[bookmarksopen,colorlinks=true,linkcolor=dark_green,citecolor=dark_red,urlcolor=dark_red,linktocpage=false]{hyperref}
\usepackage{amsmath,amssymb,cite,mathtools}
\definecolor{dark_blue}{rgb}{0,0,0.6}
\definecolor{dark_green}{rgb}{0,0.4,0}
\definecolor{dark_red}{rgb}{0.6,0,0}

\newcommand{\be}{\begin{equation}}
\newcommand{\ee}{\end{equation}}

\def\thefootnote{\fnsymbol{footnote}}

\renewcommand{\thefootnote}{$\diamondsuit$\arabic{footnote}}
\begin{document}

\begin{titlepage}
\begin{flushright}
DESY 19-102\\
\end{flushright}

\vspace{1cm}
\begin{center}
{\Large\bf\color{black}
Relativistic bubble collisions - a closer look
}\\
\bigskip\color{black}
\vspace{1cm}{
{\large Ryusuke Jinno$^a$, Thomas Konstandin$^a$, and Masahiro Takimoto$^b$}
\vspace{0.3cm}
} \\[2mm]
{\it
{$^a$ DESY, Notkestra\ss e 85, D-22607 Hamburg, Germany} \\[0.1cm]
{$^b$ Department of Particle Physics and Astrophysics, Weizmann Institute of Science,} \\[0.12cm]
{Rehovot 7610001, Israel}
}
\end{center}
\bigskip

\vspace{.4cm}

\begin{abstract}
We study scalar bubble collisions in first-order phase transitions focusing on the relativistic limit.
We propose 'trapping equation' which describes the wall behavior after collision,
and test it with numerical simulations in several setups.
We also examine the energy dynamics after collision and discuss its implications to gravitational wave production.
\end{abstract}

\bigskip

\end{titlepage}

\tableofcontents
\thispagestyle{empty}

\renewcommand{\thepage}{\arabic{page}}
\setcounter{page}{1}
\renewcommand{\thefootnote}{$\diamondsuit$\arabic{footnote}}
\setcounter{footnote}{0}

\clearpage

\section{Introduction}
\label{sec:intro}
\setcounter{equation}{0}

First-order phase transitions lead to a variety of phenomena in the early Universe 
such as baryogenesis~\cite{Kuzmin:1985mm}, 
gravitational wave (GW) production~\cite{Witten:1984rs,Hogan:1986qda,Kosowsky:1991ua,Kosowsky:1992rz,Kosowsky:1992vn,Kamionkowski:1993fg}, 
and magnetogenesis~\cite{Vachaspati:1991nm}, among others.
All these phenomena occur during the nucleation of bubbles, their expansion, and collisions.
It is therefore important to understand the bubble dynamics 
in order to predict any possible signal of a first-order phase transition.

The dynamics of bubble expansion is determined by the balance between pressure and friction to the walls:
they receive pressure from the supercooled fluid,
while friction arises from the particle species which receive masses from the change in the scalar field value.
While there are discussions that friction is much more efficient than previously thought~\cite{Bodeker:2009qy, Bodeker:2017cim},
it is still important to understand the bubble behavior in a scalar-dominated system,
since a scalar-dominated transition is likely to occur when the latent heat is much larger than the plasma energy density 
(e.g.~in near-conformal phase transitions with extreme supercooling~\cite{Randall:2006py,Espinosa:2008kw,Konstandin:2011dr,Hambye:2013sna,Jaeckel:2016jlh,Jinno:2016knw,Marzola:2017jzl,Iso:2017uuu,Chiang:2017zbz,vonHarling:2017yew,Bruggisser:2018mus,Bruggisser:2018mrt,Hambye:2018qjv,Baldes:2018emh,Hashino:2018wee,Prokopec:2018tnq,Brdar:2018num,Marzo:2018nov,Baratella:2018pxi,Fairbairn:2019xog}\footnote{
Ref.~\cite{Ellis:2019oqb} discusses the required value of $\alpha$ 
(latent heat density normalized by the plasma energy density just before the transition)
for the scalar field to be dominant in a model having this property.
Also, see Ref.~\cite{Ellis:2018mja} for the maximal value of $\alpha$ for polynomial potentials.
}).
If such a transition occurs, the energy released inside cosmological-scale bubbles accumulates mostly on the walls,
and the resulting relativistic $\gamma$ factor is huge (say $\gtrsim 10^{10}$) at the time of collisions.
Recent numerical simulations have been performed in Ref.~\cite{Child:2012qg,Braden:2014cra,Braden:2015vza,Bond:2015zfa,Cutting:2018tjt},
often with a focus on oscillons and gravitational wave production. However, simulating bubbles with a large $\gamma$ factor is on the lattice currently impossible in $3 + 1$ dimensions. 

Given this, the aim of the present work is to develop a method to understand relativistic bubble collisions analytically. 
We find that, in the relativistic limit, there is a simple governing equation that determines the wall behavior. 
In particular there is the possibility that after the collision the scalar field bounces back to the symmetric phase 
and is trapped there. This equation tells us whether the trapping at the false vacuum occurs after collisions. 
This has huge impact on the GW spectrum, since the GW spectrum takes quite different forms depending on 
whether the scalar field is damped at the collision point (in which case 
the envelope approximation~\cite{Kosowsky:1992rz,Kosowsky:1992vn,Huber:2008hg,Jinno:2016vai} should be appropriate)
or the walls pass through each other mostly unhindered 
(in which case the flow approximation~\cite{Jinno:2017fby,Konstandin:2017sat} should be appropriate). 
Therefore, our study will help to identify the GW spectrum resulting in scalar-dominated transitions.

The organization of the paper is as follows.
In Sec.~\ref{sec:criterion} we first outline our setup and introduce the governing equation, which we call 'trapping equation'.
In Sec.~\ref{sec:test} we numerically check the validity of this equation with a variety of setups.
In Sec.~\ref{sec:appl} we discuss the application of the trapping equation to $U(1)$ or $SU(2)$ breaking transitions.
In Sec.~\ref{sec:energy} we study the energy localization, which is important in determining the shape of the GW spectrum.
Sec.~\ref{sec:conc} is devoted to conclusions.

\section{The scalar dynamics after collisions}
\label{sec:criterion}
\setcounter{equation}{0}

The goal of this paper is establish an easy criterion to decide how 
the scalar field behaves after the collision of two highly-relativistic bubbles.
For most parts, we will work in the planar approximation which should be well justified 
at the first stages of the collision.

The scalar field obeys the Klein-Gordon equations 
\be
\Box \phi + \frac{dV}{d\phi} = 0 \, ,
\ee
where we neglected all interactions but the self-interactions of the scalar field that are encoded in the scalar potential $V$. A soliton connects two local minima of the potential and the form of the potential will determine the shape of the soliton while it is accelerating. 

\begin{figure}
\begin{center}
\includegraphics[width=\columnwidth]{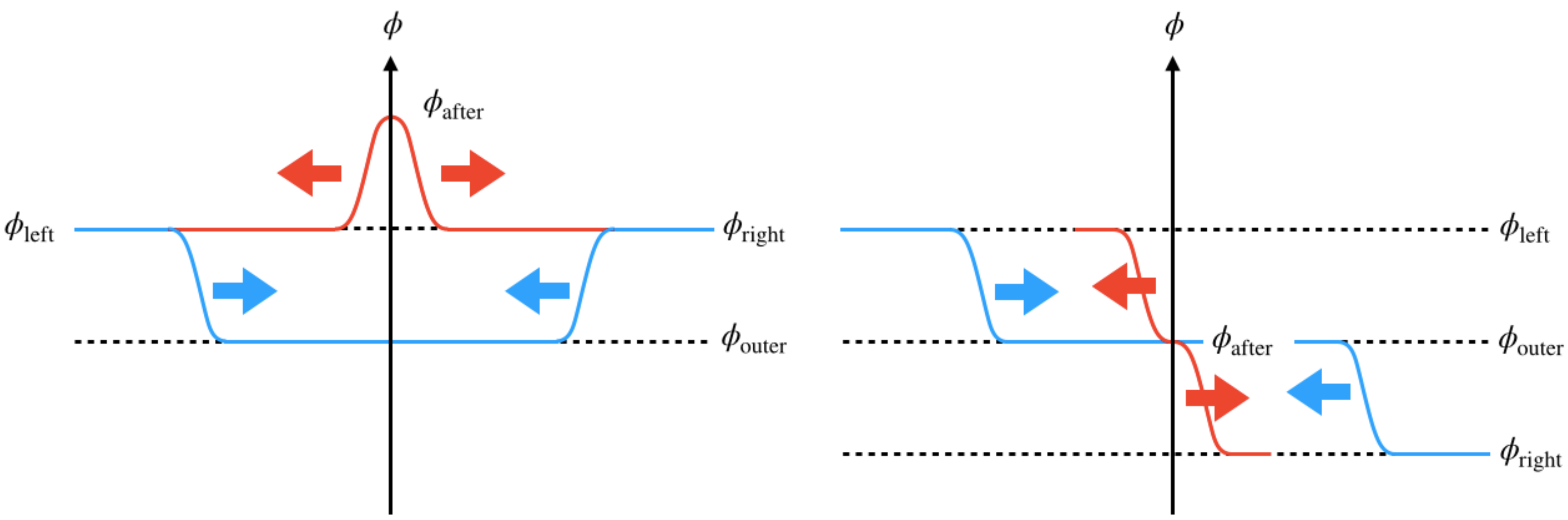}
\caption{\small
The collision of two solitons (left) and the collision of a soliton and an anti-soliton (right) in a periodic potential. The walls are shown before (blue) and after (red) the collision.
}
\label{fig:solitons}
\end{center}
\end{figure}

One important point is that in the relativistic limit, the kinetic term will dominate the dynamics and the potential is actually irrelevant during the collision. 
Since the kinetic term only leads to a linear term in the equations of motion, the superposition of two solitons will persist even after collision 
as long as the solitons collide with a highly-relativistic velocity. 
The solitons consist of an 'inner' region where the scalar field $\phi$ has the values $\phi_{\rm left}$ and $\phi_{\rm right}$. 
The solitons are expanding into an 'outer' region, where the scalar field has a value $\phi_{\rm outer}$. As long as the superposition of the solitons persists, 
the scalar field has to acquire the value 
\be
\phi_{\rm after} =  \phi_{\rm left} + \phi_{\rm right} - \phi_{\rm outer} \, ,
\ee
after the collision. This is exemplified in Fig.~\ref{fig:solitons} where we show the collision of two solitons and the collision of a soliton and an anti-soliton 
in a periodic potential (in a periodic potential $\phi_{\rm after}$ is again a minimum of the potential).  

However, in the long run and in non-periodic potentials, the scalar field will not pertain the value $\phi_{\rm after}$ 
since this is often not a local minimum of the potential. In essence, the scalar field will start rolling down the potential. 
The boundary conditions are set by the solitons flying apart which induces a $SO(1,1)$ symmetric initial condition and the solution can only depend on 
the lightfront coordinate $s = \sqrt{t^2 - x^2}$. 

\begin{figure}
\begin{center}
\includegraphics[width=0.65\columnwidth]{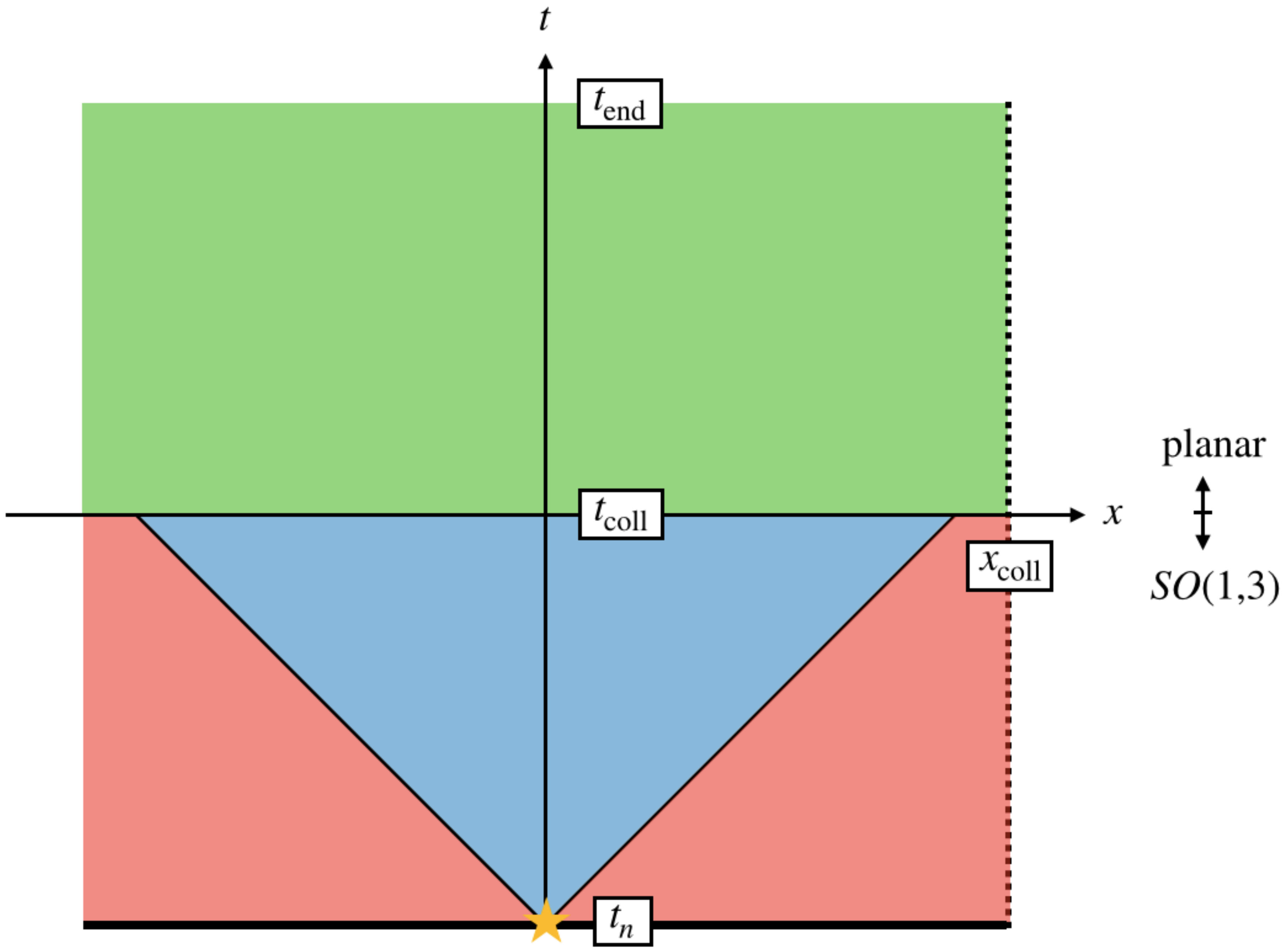}
\caption{\small
Illustration for our setup.
The bubble nucleates at $t = t_n$ at the position of the star.
The bounce configuration is located along the thick black line.
The scalar field configuration in the spacelike region from the nucleation point (red) is 
related to the bounce configuration (\ref{eq:phi_spacelike}),
while the one in the timelike (blue) region is related to (\ref{eq:phi_timelike}).
The collision time of the bubble $t_{\rm coll}$ is also indicated in the figure.
To simulate two colliding bubbles,
we impose reflecting boundary conditions at $x = x_{\rm coll}$,
which is taken to be slightly larger than $t_{\rm coll} - t_n$.
Our interest lies in the evolution of the system in the green region.
We take the simulation end $t_{\rm end}$ so that $t_{\rm end} - t_{\rm coll} \approx  x_{\rm coll}$
to guarantee that the most relativistic components around $x \simeq x_{\rm coll}$ at $t = t_{\rm coll}$
propagates back to $x \simeq 0$ at $t = t_{\rm end}$.
The expanding bubble has $SO(1,3)$ symmetry before $t = t_{\rm coll}$, 
while we approximate the system to be planar symmetric after collision to simplify the simulation.
}
\label{fig:spacetime}
\end{center}
\end{figure}

\begin{figure}
\begin{center}
\includegraphics[width=0.48\columnwidth]{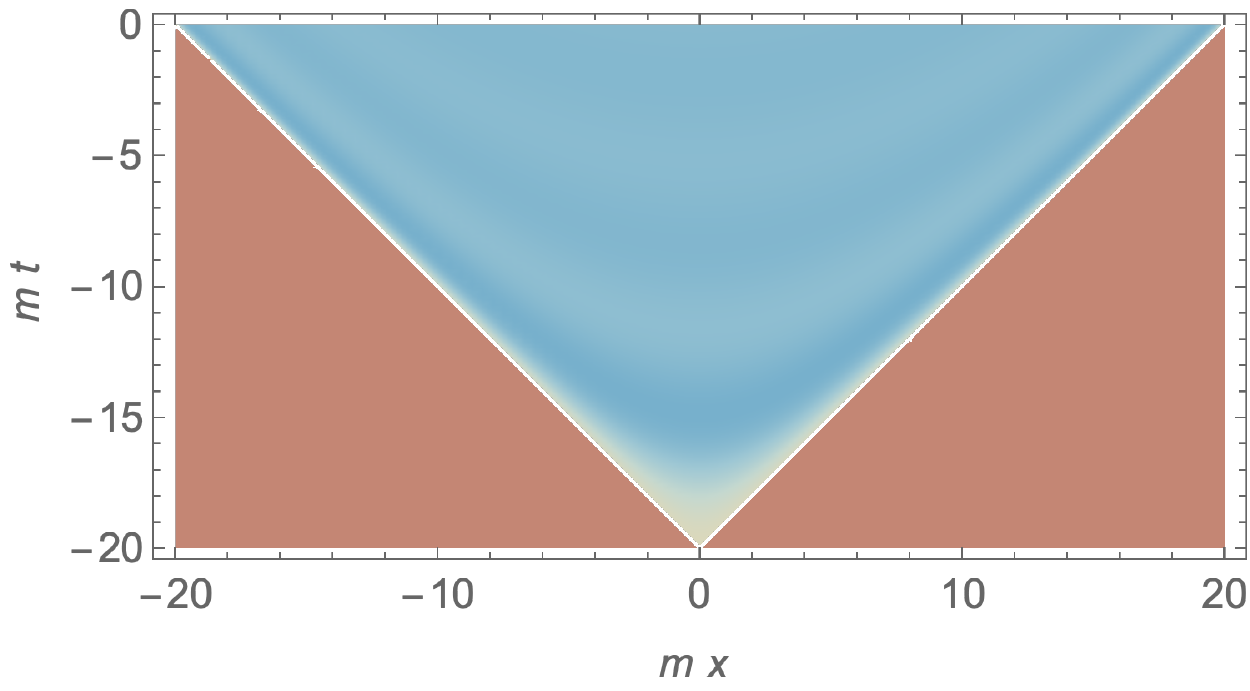}
\includegraphics[width=0.48\columnwidth]{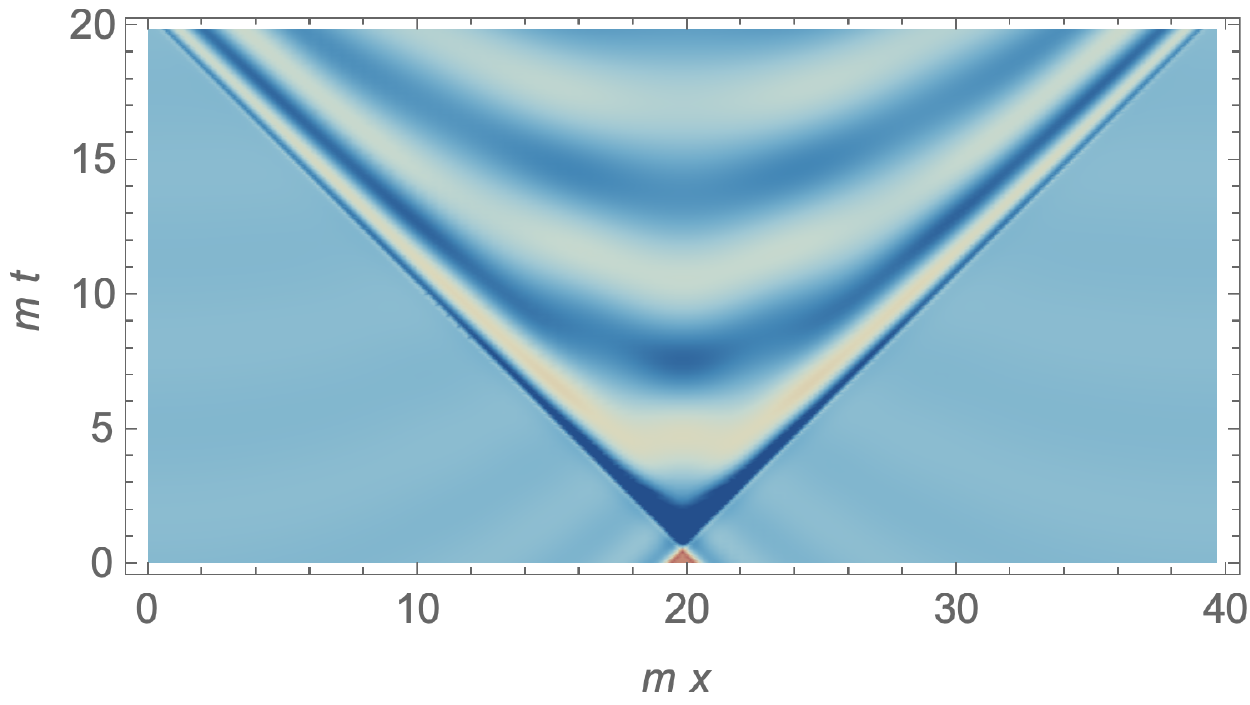}
\caption{\small
A density plot of the scalar field in a hierarchical model (see Fig.~\ref{fig:Hierarchy_V}).
Red and blue regions correspond to the false ($\phi = 0$) and true vacua ($\phi = 1$), respectively.
{\bf Left side:} Before $t_{\rm coll}$, corresponding to the red and blue regions in Fig.~\ref{fig:spacetime}.
{\bf Right side:} After $t_{\rm coll}$, corresponding to the green region in Fig.~\ref{fig:spacetime}.
}
\label{fig:spacetime_example}
\end{center}
\end{figure}

If the potential height at the $\phi$ value after settling down differs from $V(\phi_{\rm left})$ or $V(\phi_{\rm right})$, 
the corresponding wall can decelerate/accelerate which breaks the $SO(1,1)$ symmetry. 
This will have a large impact on the gravitational wave spectrum created. 
For almost degenerate potential values, the walls will only use energy through the expansion and 
the model proposed in Refs.~\cite{Jinno:2017fby,Konstandin:2017sat} is likely to describe the GW production. 
If there is a large potential difference, the wall is quickly decelerated and the envelope approximation is 
more likely to describe the GW production~\cite{Kosowsky:1992rz,Kosowsky:1992vn,Huber:2008hg,Jinno:2016vai}.
However, on times scales much smaller than the bubble separation, these effects are quite small and 
the $SO(1,1)$ symmetry is seen in most of our simulations (see Figs.~\ref{fig:spacetime} and \ref{fig:spacetime_example}).

Under these assumptions the dynamics of the scalar field close to the collision point is governed by the 'trapping equation'
\be
\partial_s^2 \phi + \frac{1}{s}\partial_s \phi + \frac{dV}{d\phi}
= 0 \, .
\label{eq:trapping}
\ee
Here, the coordinate $s$ is the $SO(1,1)$ radial direction with the collision point in the origin, $s = \sqrt{t^2 - x^2}$. 
This equation can be easily solved numerically, which is what we will do in the next section in comparison to scalar field simulations to test our hypothesis.

\section{Testing the trapping equation}
\label{sec:test}
\setcounter{equation}{0}

\subsection{Setup and initial conditions}
\label{subsec:Setup}

In order to test the trapping equation, 
we will assume planar symmetry for the colliding bubble walls. 
For the scalar configuration just before collision, we use initial conditions that are derived from the $3 + 1$ dimensional setup (see Fig.~\ref{fig:spacetime}).
Notice that in Secs.~\ref{subsec:Z2} and \ref{subsec:Z2mod} the potential minima are degenerate 
and the $3 + 1$ solutions become the exact soliton profiles in $1 + 1$ dimensions. 

The scalar field fulfills the equations of motion
\be
(\partial_t^2 - \partial_x^2) \phi + \frac{dV}{d\phi} = 0,
\ee
after collision, which assumes planar symmetry. 
Another important quantity to track is the energy density, which is given by 
\be
\rho = \frac{1}{2}(\partial_t \phi)^2 + \frac{1}{2}(\partial_x \phi)^2 + V(\phi).
\ee
This will be the relevant indicator how gravitational waves are produced in the present scenario. 
A more detailed discussion of this topic will be given in section \ref{sec:energy}.

The next question concerns the initial conditions of the scalar field before collision, i.e.~the shape of the soliton during acceleration. 
Even though our actual simulation is only $1 + 1$ dimensional, 
we mainly use the $3 + 1$ dimensional shape of the soliton. 
These two configurations can be sizable different due to the friction term in the bounce equation. 
The evolution of the single bubble configuration after nucleation at $t = t_n$ is as follows:
\begin{itemize}
\item
In the spacelike region (red) the scalar configuration is related to the bounce configuration $\bar{\phi}$ 
through $SO(1,3)$ symmetry. That is,
\be
\phi(t,r)
= 
\bar{\phi} \left( \sqrt{- (t - t_n)^2 + x^2} \right),
\ee
where $\bar{\phi}(s = \sqrt{- (t - t_n)^2 + x^2})$ satisfies the bounce equation of motion:
\be
\frac{d^2 \bar{\phi}}{ds^2}
+ \frac{3}{s} \frac{d\bar{\phi}}{ds}
- \frac{dV}{d\bar{\phi}}
= 
0.
\label{eq:phi_spacelike}
\ee
\item
In the timelike region (blue) the scalar configuration is again related through $SO(1,3)$ symmetry 
\be
\phi(t,x)
= 
\tilde{\phi} \left( \sqrt{(t - t_n)^2 - x^2} \right),
\ee
to the solution $\tilde{\phi}(s = \sqrt{(t - t_n)^2 - x^2})$ of the following equation of motion:
\be
\frac{d^2 \tilde{\phi}}{ds^2}
+ \frac{3}{s} \frac{d\tilde{\phi}}{ds}
+ \frac{dV}{d\tilde{\phi}}
= 
0.
\label{eq:phi_timelike}
\ee
In particular, the field performs oscillations around the new local minimum of the potential.
The extent of these oscillations is quite different in $3 + 1$ dimensions compared to $1 + 1$ dimensions 
and we use the former as mentioned before even though the simulation is lower dimensional.
\end{itemize}
In the next subsections we will discuss a variety of models and test the trapping equation (\ref{eq:trapping}).
First, we will discuss potentials with degenerate minima and a sizable barrier. 
In this case, trapping in the old phase is very likely. 
Next, we modify the potential beyond the new phase (but keep degenerate minima) and study for which parameters the simulation 
and the trapping equation predict a bounce back into the old phase. 
This allows for a quantitative test of the trapping equation. 
Then the opposite case is studied: potentials with a large hierarchy between the two phases and small barriers. 
First we study the idealized case of an infinitesimal barrier (that only enters in the initial conditions) and then move to more realistic models 
with and without a $Z_2$ symmetry.
In the following numerical simulations we use the time discretization $\Delta t = 0.1 \Delta x$
except for Figs.~\ref{fig:Quartic_phi}, \ref{fig:Quartic_trapped}, \ref{fig:Quartic_escaped}, and \ref{fig:Quartic_escaped_limit}, 
in which we use $\Delta t = 0.05 \Delta x$.
The spatial discretization $\Delta x$ is chosen depending on the setup.

\subsection{Toy model 1: Simple $Z_2$ potential}
\label{subsec:Z2}

\begin{figure}
\begin{center}
\includegraphics[width=0.45\columnwidth]{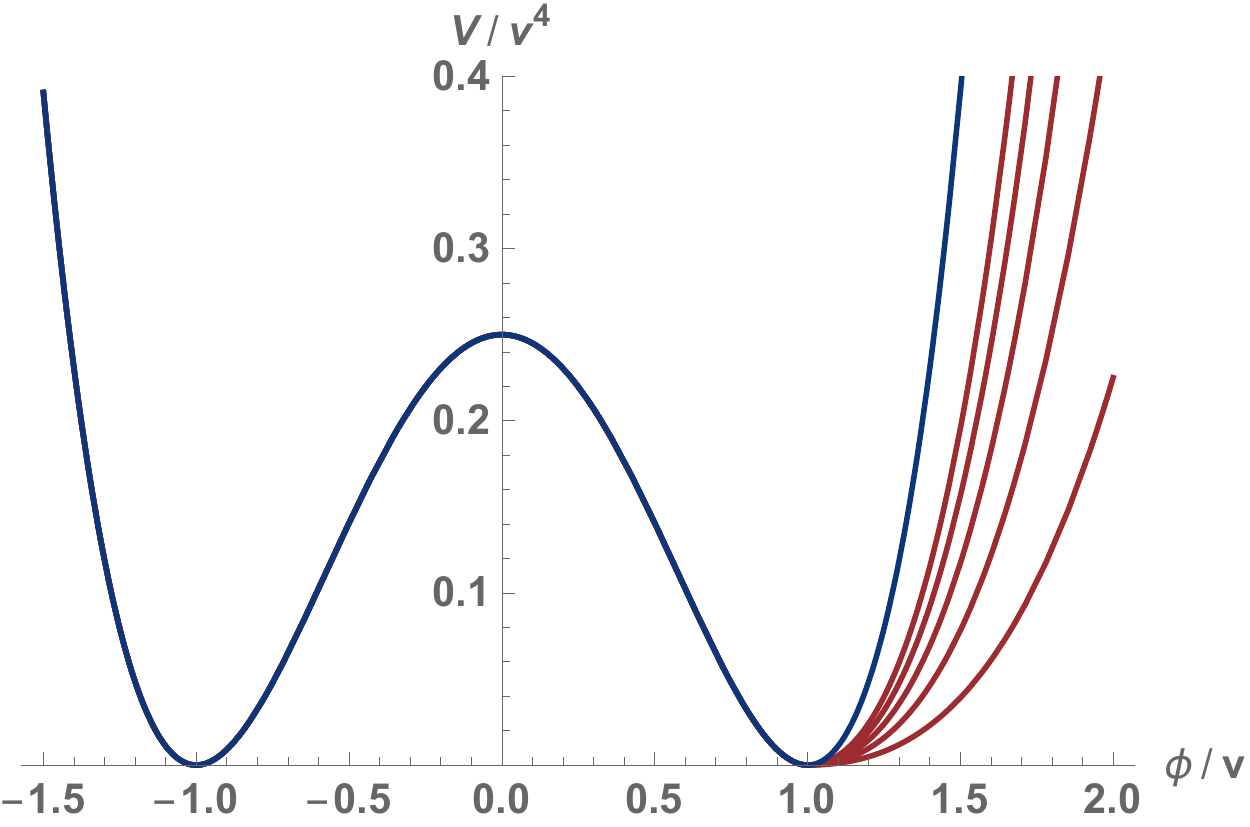}
\caption{\small
Potential shape of the simple and modified $Z_2$ model. 
The top line shows the $Z_2$-symmetric potential discussed in Sec.~\ref{subsec:Z2}, 
while the other lines belong to the modified potential in Sec.~\ref{subsec:Z2mod}
with $\lambda = 0.1, 0.2, 0.3, 0.4$, and $0.5$.
}
\label{fig:Z2_V}
\end{center}
\vskip 0.5cm
\begin{center}
\includegraphics[width=0.48\columnwidth]{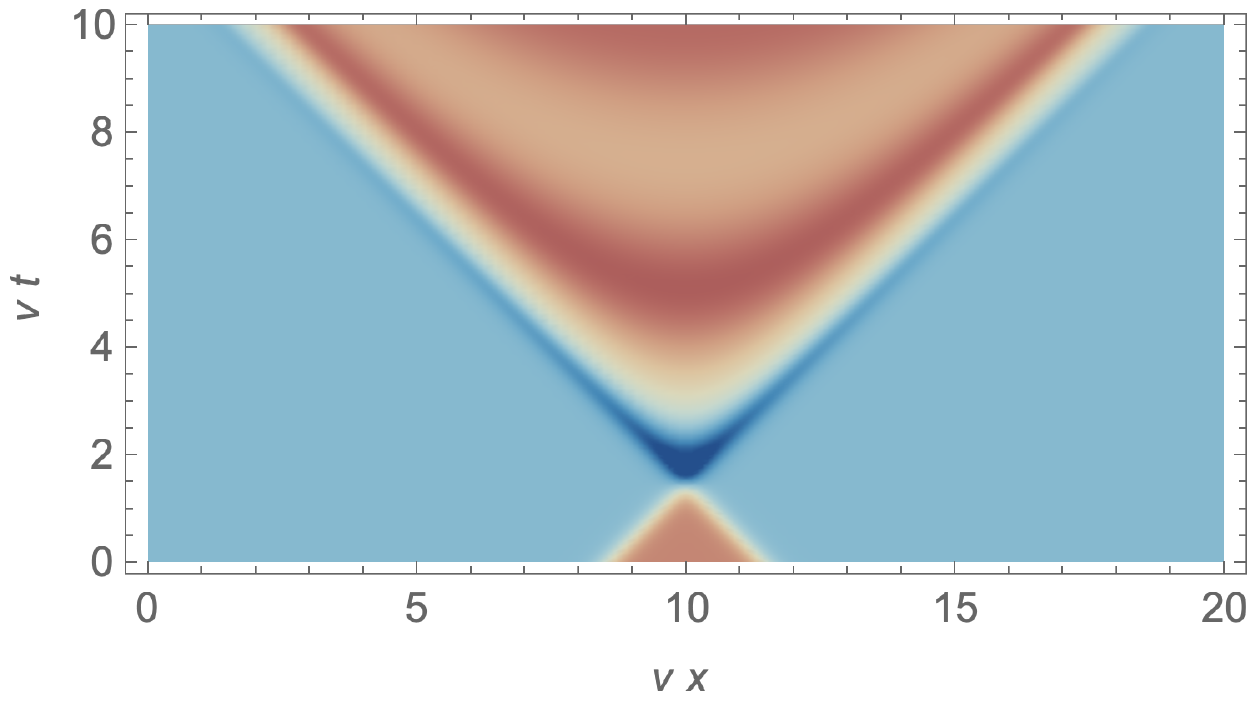}
\includegraphics[width=0.48\columnwidth]{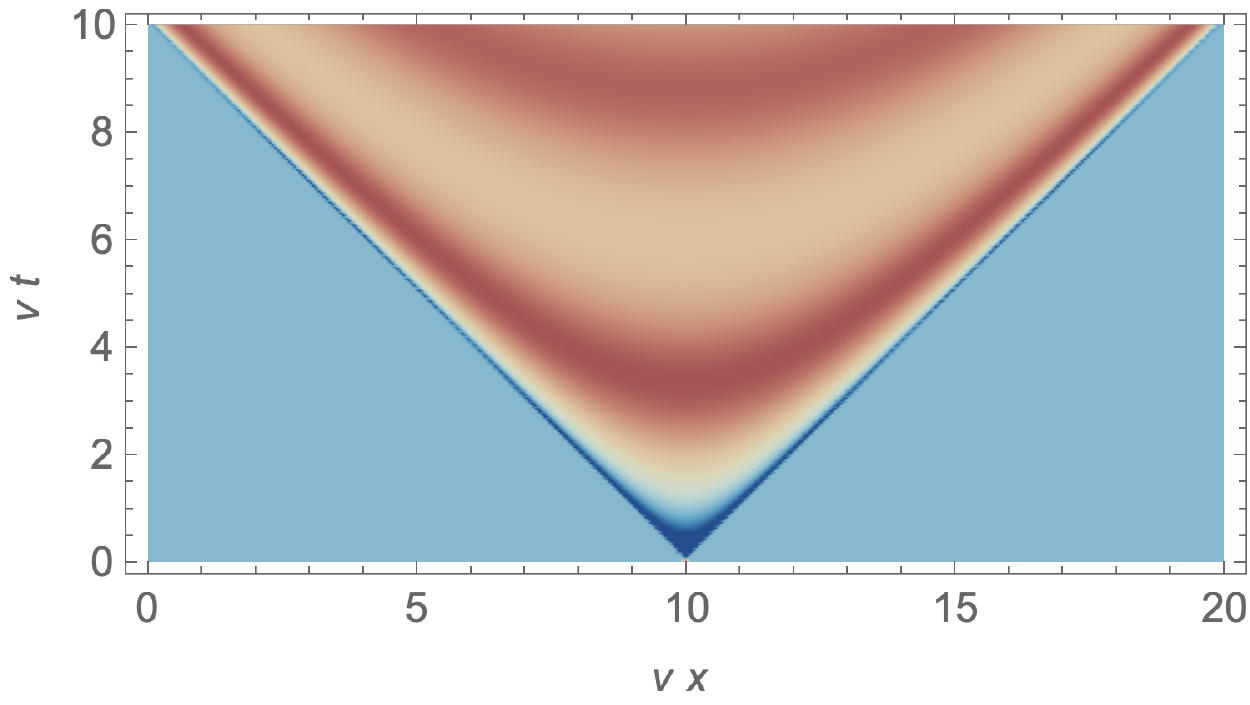}
\caption{\small
Simple $Z_2$ potential. Profile of $\phi$ for the $Z_2$ potential with a $\gamma$ factor of $5$ (left) and $100$ (right).
The blue and red regions indicate the positive and negative vacua, respectively.
}
\label{fig:Z2_phi}
\end{center}
\end{figure}

We first consider a $Z_2$-symmetric degenerate potential
\be
V = \frac{1}{4}(\phi^2 - v^2)^2.
\ee
As mentioned in Sec.~\ref{subsec:Setup}, 
the $3 + 1$ dimensional solutions reduce to the exact soliton profiles in $1 + 1$ dimensions. 
The profile is given by
\be
\phi(t,x)
=
\pm \, v \, \tanh 
\left[
\frac{\gamma}{\sqrt{2}}
\left(
x - \sqrt{1 - \frac{1}{\gamma^2}} t
\right)
+ \delta
\right],
\label{eq:tanh}
\ee
with $\gamma$ being the relativistic $\gamma$ factor.
We initially prepare soliton states according to (\ref{eq:tanh}) with the value $-v$ in the outer region and evolve the system with  reflecting boundary conditions at 
$x = x_{\rm coll} = 10/v$.
According to the trapping equation (\ref{eq:trapping}), 
the scalar field bounces back to the minimum at $\phi = -v$ after collision in the large $\gamma$ limit.
We show the time evolution for $\gamma = 5$ (small $\gamma$) and $100$ (large $\gamma$) in Fig.~\ref{fig:Z2_phi}.
We evolve the system from $t = t_{\rm coll} = 0$ to $t = t_{\rm end} = 10/v$.
We take $400 \gamma$ points for $0 < x < x_{\rm coll}$ so that $\Delta x = 1/40\gamma/v$ for the spatial discretization, 
and we choose the phase $\delta$ so that the argument inside the $\tanh$ becomes $5$ 
at the boundary $x = x_{\rm coll}$ at $t = t_{\rm coll}$.
It is seen that, for both large and small $\gamma$, the scalar field bounces back to $\phi = -v$.
This is consistent with the trapping equation (\ref{eq:trapping}).

\subsection{Toy model 2: Modified $Z_2$ potential}
\label{subsec:Z2mod}

\begin{figure}
\begin{center}
\includegraphics[width=0.48\columnwidth]{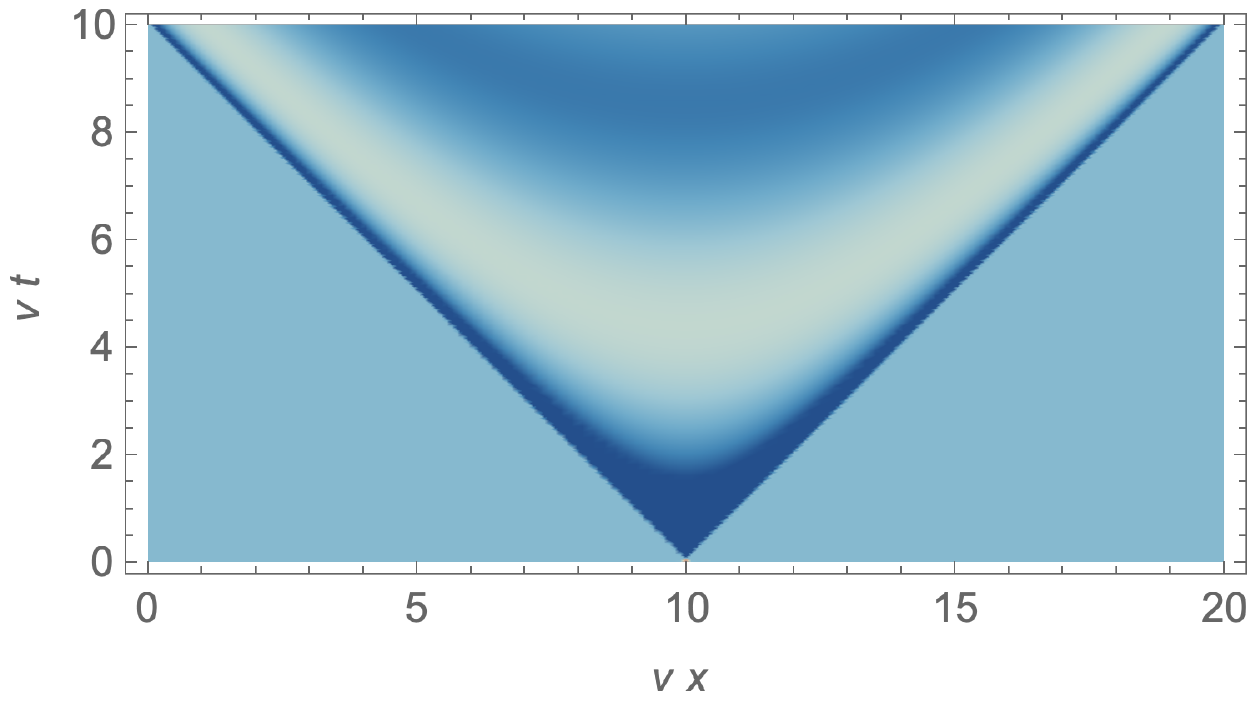}
\includegraphics[width=0.48\columnwidth]{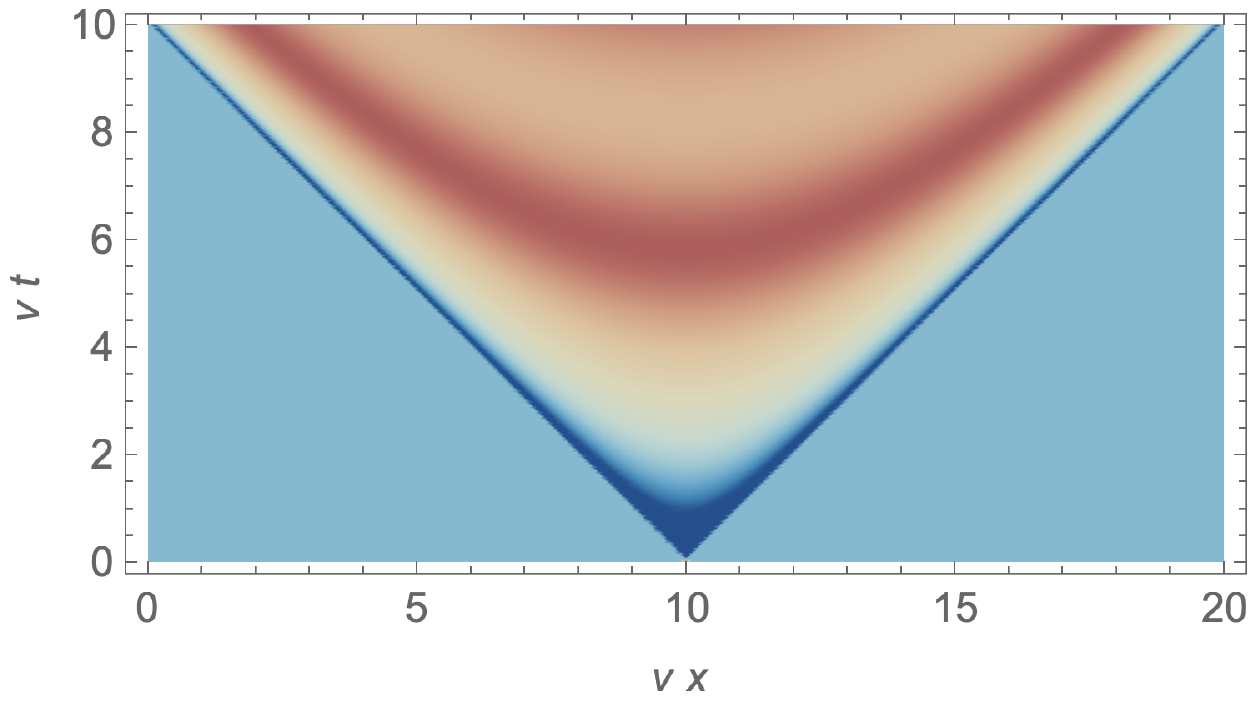}
\caption{\small
Profile of $\phi$ for the modified $Z_2$ potential with a $\gamma$ factor of $100$ 
with $\lambda = 0.1$ ($< \lambda_{\rm th} \simeq 0.186$, left) and $\lambda = 0.3$ ($> \lambda_{\rm th}$, right).
The blue and red regions indicate the positive and negative vacua, respectively.
For the former the scalar field does not bounce back to the vacuum at $\phi = -v$ while it does for the latter.
Compare with the right panel of Fig.~\ref{fig:Z2_phi}.
\label{fig:Z2mod_phi}
}
\end{center}
\vskip 0.7cm
\begin{center}
\includegraphics[width=0.6\columnwidth]{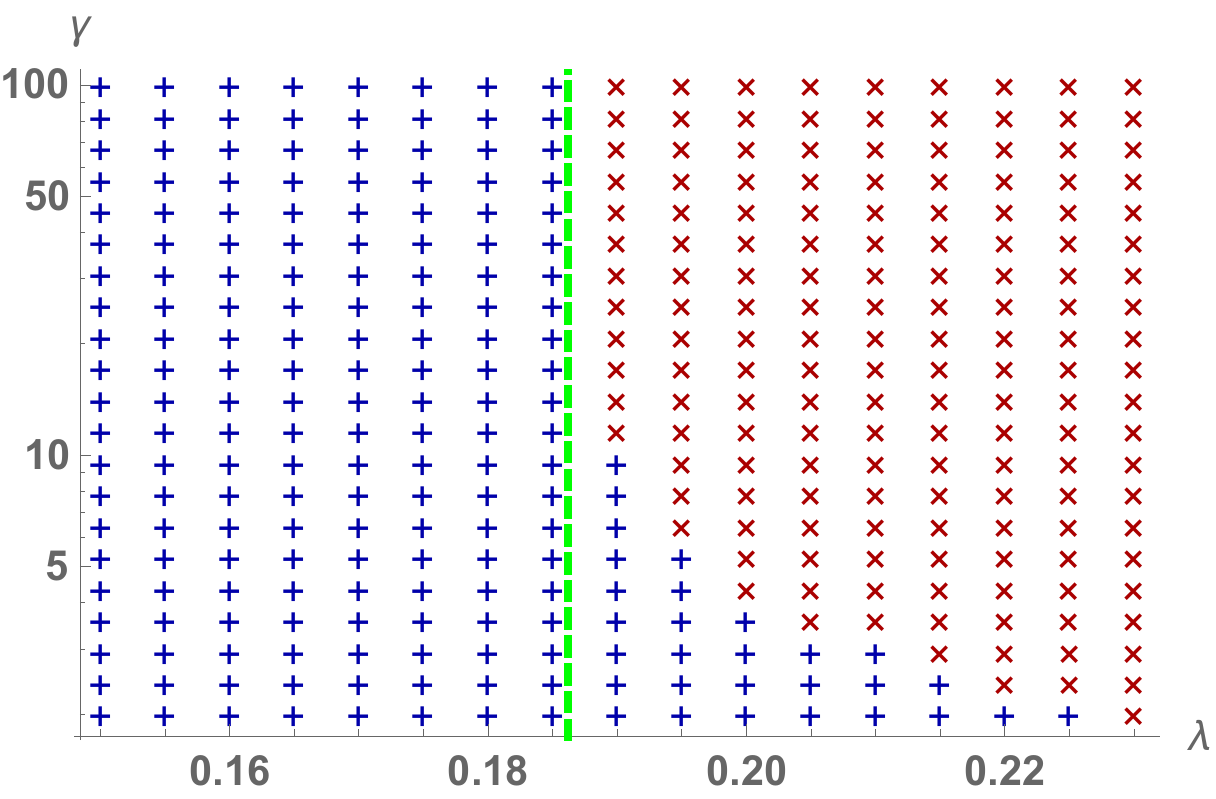}
\caption{\small
Modified $Z_2$ potential. The red and blue points are parameter points where 
$\phi (t = t_{\rm end}) \lessgtr 0$, respectively.
The green-dashed line is $\lambda = \lambda_{\rm th} \simeq 0.186$ predicted by Eq.~(\ref{eq:trapping}).
}
\label{fig:Z2mod_lambda_gamma}
\end{center}
\end{figure}

We next modify the $Z_2$-symmetric potential slightly to test the validity of the trapping equation quantitatively:
\be
V =
\left\{
\begin{matrix}
\displaystyle
\frac{1}{4}(\phi^2 - v^2)^2
&
~~
(\phi \leq v),
\\[0.3cm]
\displaystyle
\frac{\lambda}{4}(\phi^2 - v^2)^2
&
~~
(\phi > v).
\end{matrix}
\right.
\ee
Here $\lambda$ is a free parameter which controls the steepness for $\phi > v$ (see Fig.~\ref{fig:Z2_V}).
We solve the same evolution equation as before with the same initial profile.
Note that the potential modification for $\phi > v$ does not change the soliton profile.
The motivation for making the potential shallower beyond the broken phase is 
to inhibit that the field is driven back to the symmetric phase after collision.
According to the trapping equation (\ref{eq:trapping}),
$\phi$ settles down to the positive minimum for $\lambda < \lambda_{\rm th} \simeq 0.186$,
while it bounces back to the negative minimum for $\lambda > \lambda_{\rm th}$,
where $\lambda_{\rm th}$ is a threshold value.
Below we will see that this indeed holds in the large $\gamma$ limit.
In Fig.~\ref{fig:Z2mod_phi}, we show the profile of $\phi$ with $\lambda = 0.1 < \lambda_{\rm th} \simeq 0.186$ (left) 
and $\lambda = 0.3 > \lambda_{\rm th}$ (right).
The blue and red regions indicate that $\phi$ is in the positive and negative vacua, respectively.
We evolve the system from $t = t_{\rm coll} = 0$ to $t = t_{\rm end} = 10/v$
with $200 \gamma$ points for $0 < x < x_{\rm coll}$.
In contrast to Fig.~\ref{fig:Z2_phi},
$\phi$ settles down to the positive minimum for $\lambda = 0.1$, 
while $\phi$ bounces back to the negative minimum for $\lambda = 0.3$.

The parametric dependence on the parameters $\lambda$  and $\gamma$ is shown in Fig.~\ref{fig:Z2mod_lambda_gamma}. 
The red points denote parameters with a bounce back into the negative vacuum ($\phi (t = t_{\rm end}) < 0$), 
while the blue points denote parameters where $\phi$ stays in the positive vacuum ($\phi (t = t_{\rm end}) > 0$).
The green-dashed line is $\lambda_{\rm th}$ as derived from the trapping equation.
The boundary between the blue and red regions coincides very well with $\lambda_{\rm th}$ in the large $\gamma$ limit.

\subsection{Toy model 3: Hierarchical potential}
\label{subsec:Quadratic}

Next let us consider the opposite case to the previous one.
We consider a potential where the false vacuum is located at $\phi = 0$ with an infinitesimally small trap.
The true vacuum is located at $\phi = v$, and the potential shape around it is quadratic. Thus 
\be
V \simeq \frac{1}{2} m^2 (\phi - v)^2,
\ee
up to small effects very close to the symmetric phase.
As mentioned in Sec.~\ref{subsec:Setup}, 
we regard the system to be $3 + 1$ dimensional until collision (i.e. spherical),
while approximate it to be $1 + 1$ (i.e. planar) when we calculate collision dynamics.

For a quadratic potential, the initial conditions (\ref{eq:phi_spacelike}) and (\ref{eq:phi_timelike})  are solved analytically
\be
\tilde{\phi}(s)
= v
\left[
1 - \frac{2J_1(ms)}{ms}
\right],
\ee
where $J_1$ is the Bessel function.
Therefore, the scalar configuration is given by
\be
\phi(t,x)
= 
v\left[
1 - \frac{2J_1\left( m \sqrt{(t - t_n)^2 - x^2}\right)}{m \sqrt{(t - t_n)^2 - x^2}}
\right].
\ee
For the collision, we again approximate the walls to be planar.
Since the effect of the configuration in the spacelike region is infinitesimally small, 
we may approximate the configuration at the beginning of the simulation ($t = t_{\rm coll}$) as
\be
\left.
\phi(t,x)
\right|_{t = t_{\rm coll}}
\simeq
\left\{
\begin{matrix}
\displaystyle
v \left[
1 - \frac{2J_1\left( m \sqrt{(t_{\rm coll} - t_n)^2 - x^2}\right)}{m \sqrt{(t_{\rm coll} - t_n)^2 - x^2}}
\right]
~~~~
(0 < x < t_{\rm coll} - t_n),
\\[0.7cm]
0
~~~~
(t_{\rm coll} - t_n < x),
\end{matrix}
\right.
\label{eq:phiini}
\ee
and also the time derivative is given by
\be
\left.
\partial_t \phi(t,x)
\right|_{t = t_{\rm coll}}
\simeq
\left\{
\begin{matrix}
\displaystyle
\left.
v \, \partial_t
\left[
1 - \frac{2J_1\left( m \sqrt{(t_{\rm coll} - t_n)^2 - x^2}\right)}{m \sqrt{(t_{\rm coll} - t_n)^2 - x^2}}
\right]
\right|_{t = t_{\rm coll}}
~~~~
(0 < x < t_{\rm coll} - t_n),
\\[0.7cm]
0
~~~~
(t_{\rm coll} - t_n < x).
\end{matrix}
\right.
\label{eq:phiprini}
\ee
We then study the time evolution of the system with these initial conditions\footnote{
The details of the setup are as follows.
The system size is taken to be $x_{\rm coll} \equiv (t_{\rm coll} - t_n) + 5/\gamma/m$,
and we evolve the system from $t = t_{\rm coll}$ to $t = t_{\rm coll} + x_{\rm coll} \equiv t_{\rm end}$ 
with reflecting boundary conditions at $x = x_{\rm coll}$. 
The scalar configuration is taken as
\be
\phi(t,x)
=
\left\{
\begin{matrix}
\displaystyle
v
\left[
1 - \frac{2J_1\left( m \sqrt{(t - t_n)^2 - x^2}\right)}{m \sqrt{(t - t_n)^2 - x^2}}
\right]
~~~~
(0 < x < t - t_n),
\\
\displaystyle
v \, e^{-m^2 \gamma^2 (x - (t - t_n))^2}
\left[
- 1 + \frac{2J_1\left( m \sqrt{(t - t_n)^2 - (2(t - t_n) - x)^2}\right)}{m \sqrt{(t - t_n)^2 - (2(t - t_n) - x)^2}}
\right]
~~~~
(t - t_n < x),
\end{matrix}
\right.
\nonumber
\ee
in order to ensure $\left. \partial_t \phi(t,x) \right|_{t = t_{\rm coll}, x = x_{\rm coll}} \simeq 0$.
The initial conditions are given by $\left. \phi(t,x) \right|_{t = t_{\rm coll}}$ 
and $\left. \partial_t \phi(t,x) \right|_{t = t_{\rm coll}}$.
}.

\begin{figure}
\begin{center}
\includegraphics[width=0.6\columnwidth]{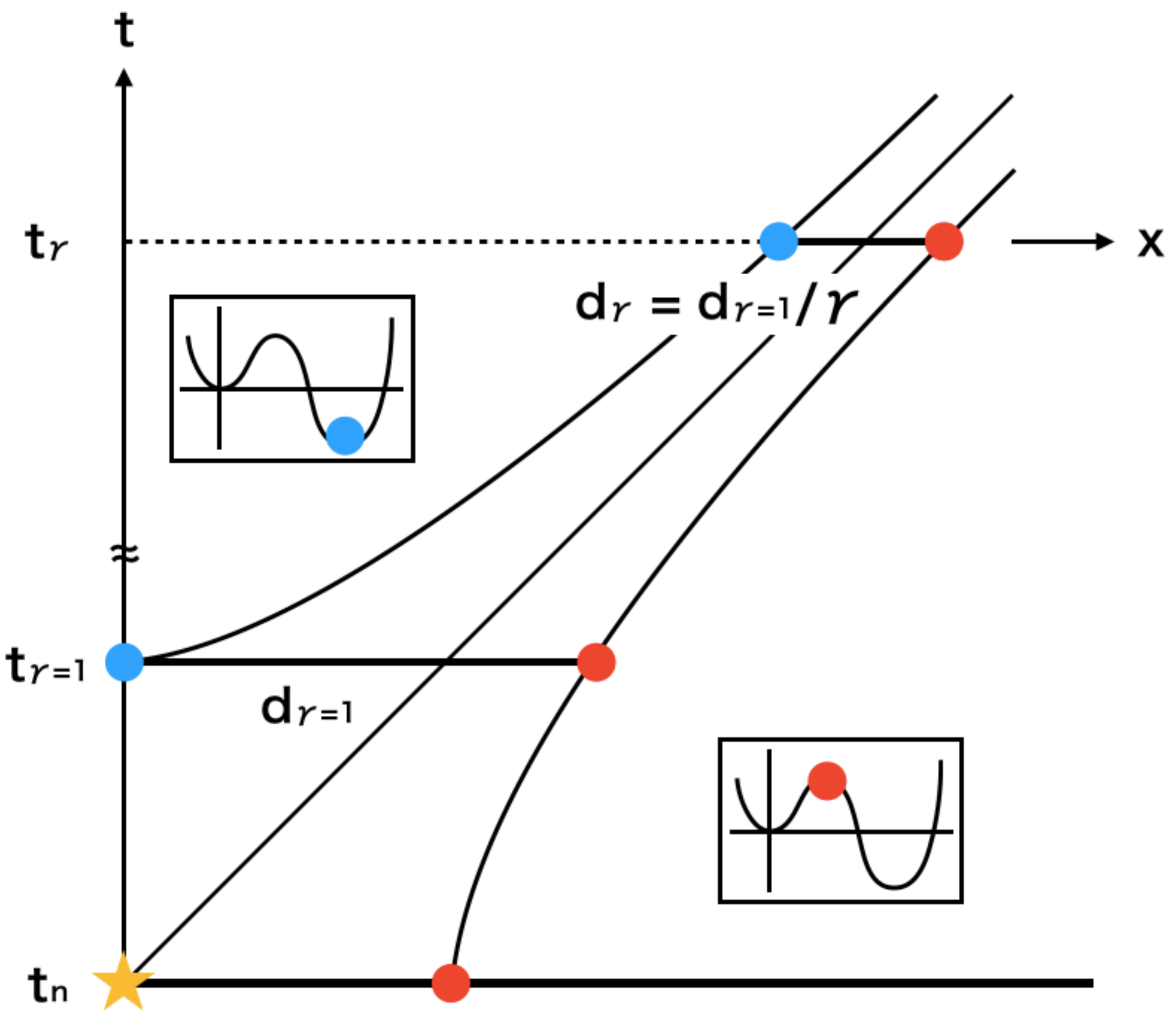}
\caption{\small
Definition of the wall $\gamma$ factor in this paper.
}
\label{fig:gamma_def}
\end{center}
\end{figure}

Since no analytic profile for the initial scalar field is known, we define the $\gamma$ factor of the colliding bubble walls 
using the Lorentz contraction of the wall (see Fig.~\ref{fig:gamma_def}):
At the nucleation time, we have the bounce configuration along $x$ direction (the thick black line), 
which has not yet reached the potential minimum.
The scalar field reaches the minimum after some time ($\sim {\rm (typical~potential~mass~scale)}^{-1}$), which we denote $t_{\gamma = 1}$.
We define $d_{\gamma = 1}$ as the spatial distance between the two points 
where the scalar field takes the minimum and the maximum at this time slice.
We define $d_\gamma$ as the distance between such spatial points at the time slice $t = t_\gamma$.
Then we define the $\gamma$ factor as the ratio of the two distances:
\be
\gamma
=
\frac{d_{\gamma = 1}}{d_\gamma}.
\ee
For the hierarchical potential we numerically find $t_{\gamma = 1} - t_n \simeq 3.83/m$.
In the simulation, we identify $t_{\rm coll}$ as $t_\gamma$ for a given value of $\gamma$.
For the spatial discretization we use $100\gamma^2$ points for $0 < x < x_{\rm coll}$.

\begin{figure}
\begin{center}
\includegraphics[width=0.45\columnwidth]{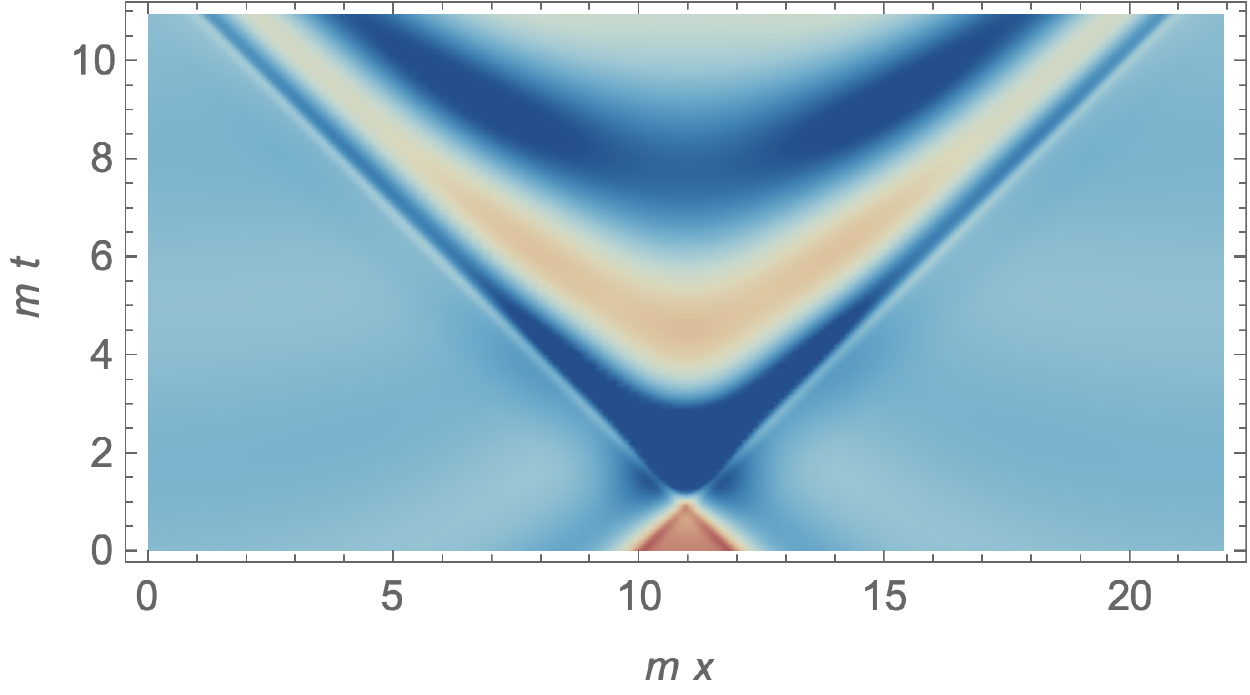}
\hskip 0.5cm
\includegraphics[width=0.4\columnwidth]{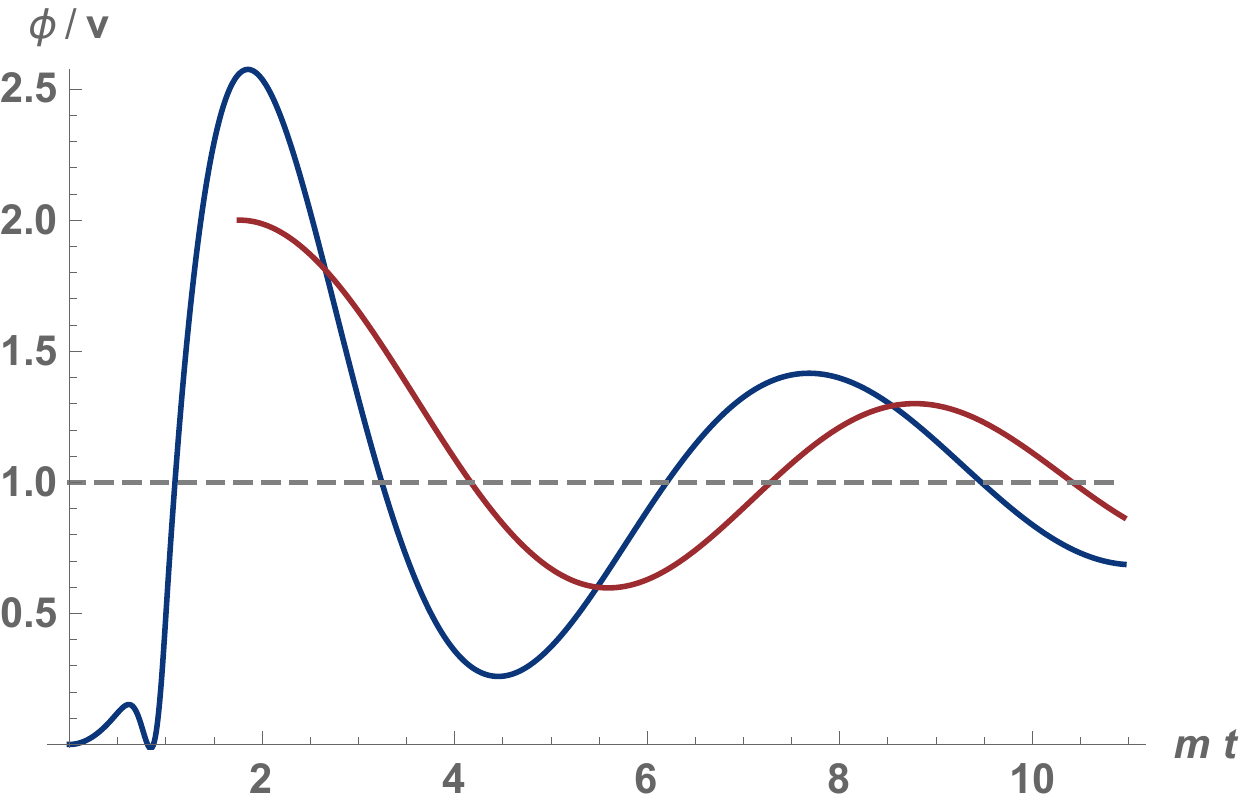}
\vskip 0.5cm
\includegraphics[width=0.45\columnwidth]{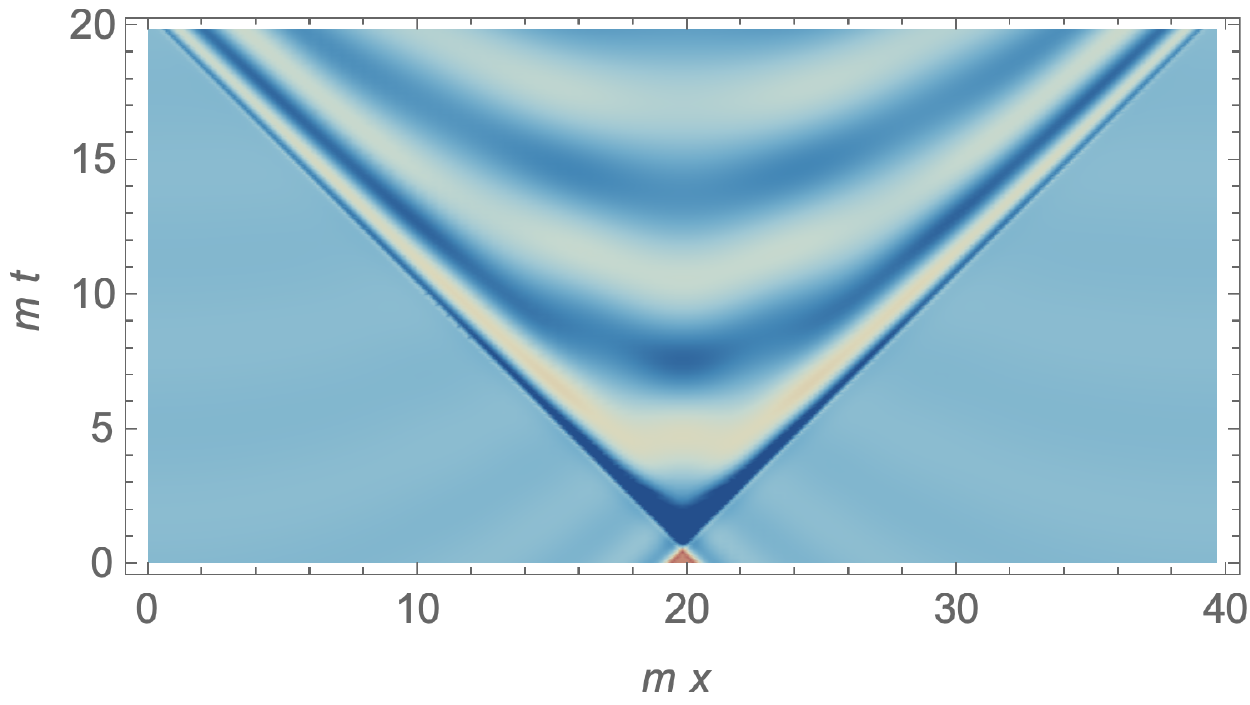}
\hskip 0.5cm
\includegraphics[width=0.4\columnwidth]{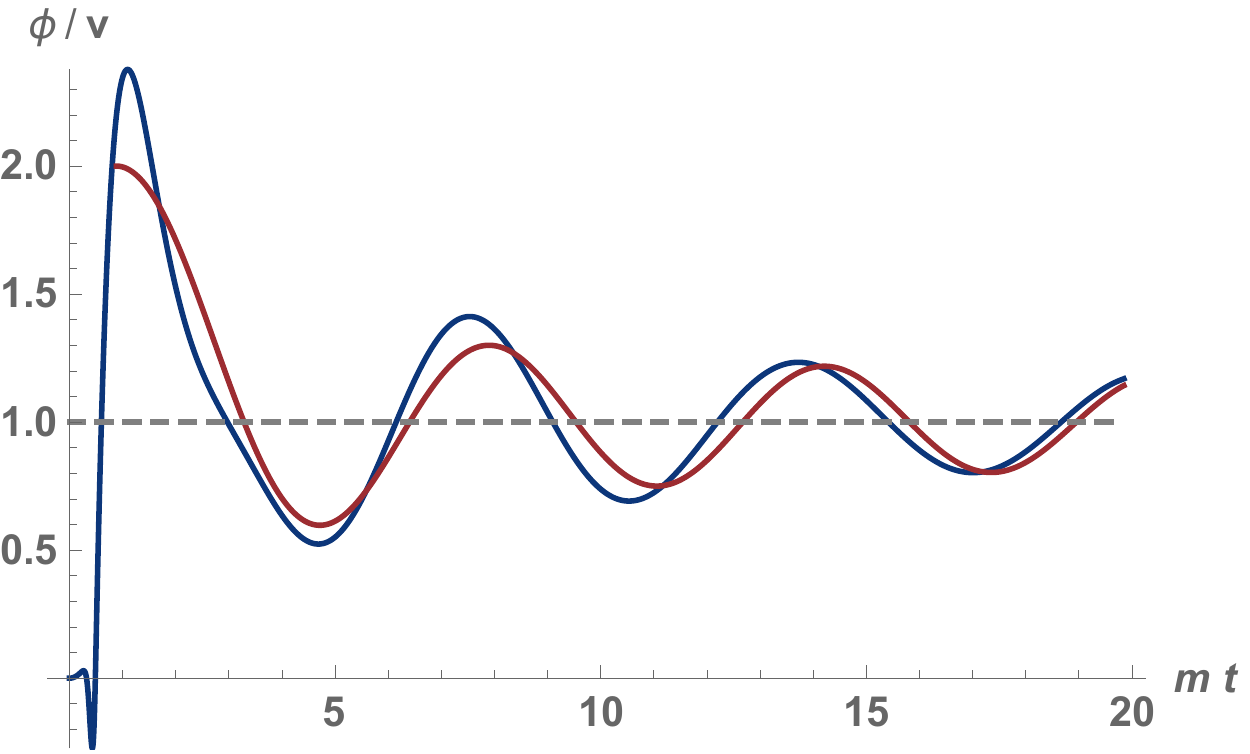}
\vskip 0.5cm
\includegraphics[width=0.45\columnwidth]{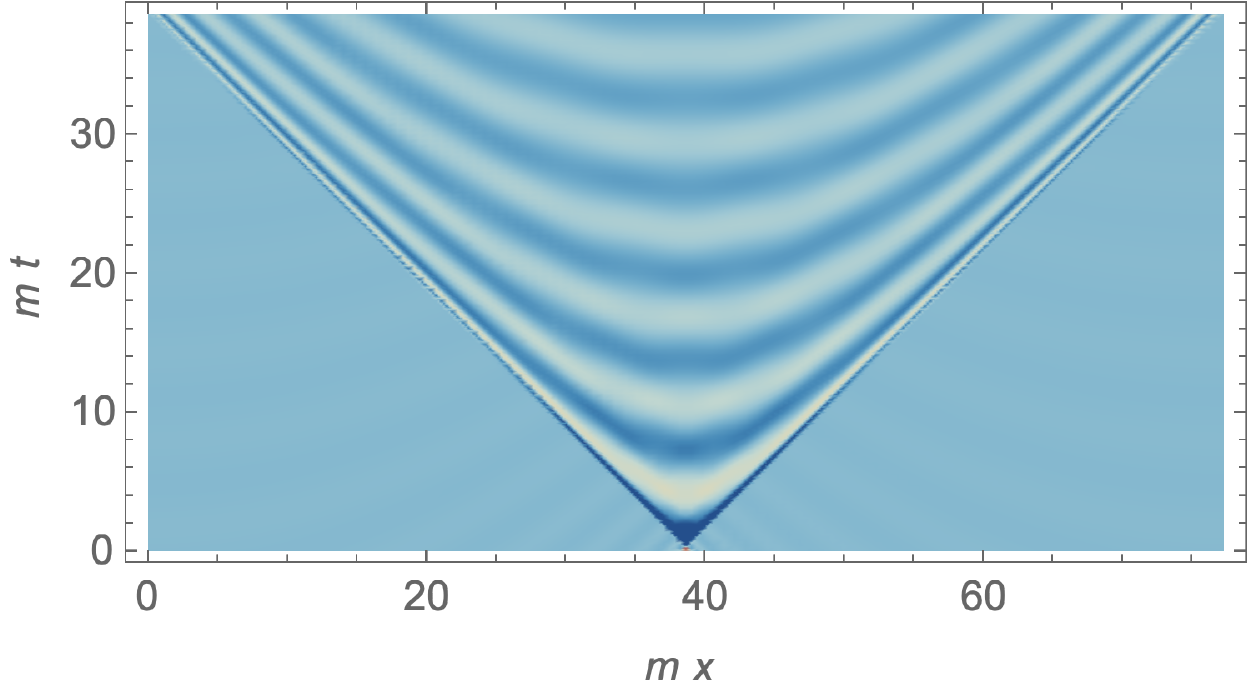}
\hskip 0.5cm
\includegraphics[width=0.4\columnwidth]{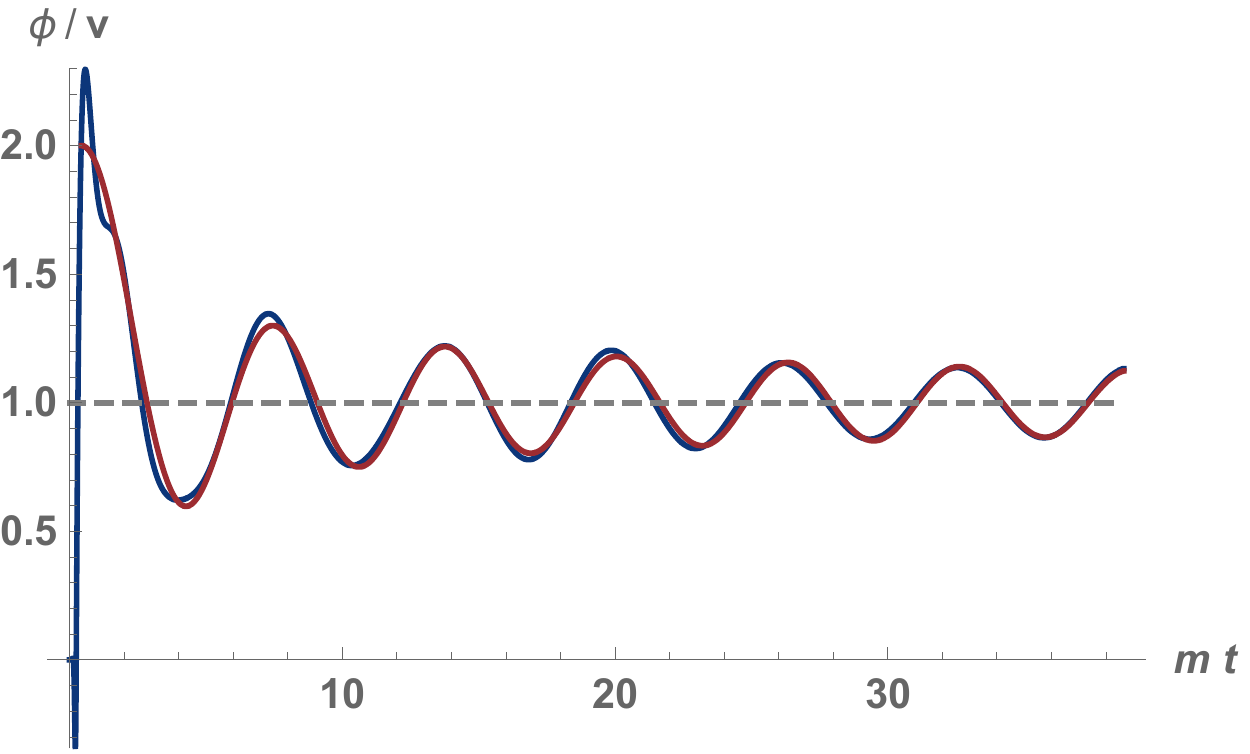}
\caption{\small
Hierarchical potential.
{\bf Left side:} 
Time evolution of $\phi$ for $\gamma = 5,10$, and $20$ from top to bottom.
{\bf Right side:} 
Time evolution of $\phi(t,x = x_{\rm coll})$ for $\gamma = 5,10$, and $20$ from top to bottom.
(Blue) Numerical evolution. (Red) Prediction from Eq.~(\ref{eq:trapping}).
}
\label{fig:Hierarchy_phi}
\end{center}
\end{figure}

Fig.~\ref{fig:Hierarchy_phi} displays the time evolution for $\gamma = 5$, $10$, and $20$ from top to bottom, respectively. 
The left panels are the value of $\phi$,
while the right panels are the time evolution at the collision point $\phi(t, x = x_{\rm coll})$. 
In the right panels the blue lines are the actual evolution in our simulation, 
while the red lines are the prediction from the trapping equation,  Eq.~(\ref{eq:trapping})
(which reduces to $\phi \simeq v \left[ 1 + J_0(m(t - t_{\rm offset})) \right]$ with $t_{\rm offset} = (3.83 + 5)/\gamma/m$).
In the offset, the contribution $3.83/\gamma/m$ comes from $t_{\gamma}$ as seen in Fig.~\ref{fig:gamma_def} 
and the contribution $5/\gamma/m$ comes from the offset between the collision and the initial time. 
We see that the prediction nicely matches the actual evolution for large $\gamma$. 

\begin{figure}
\begin{center}
\includegraphics[width=0.5\columnwidth]{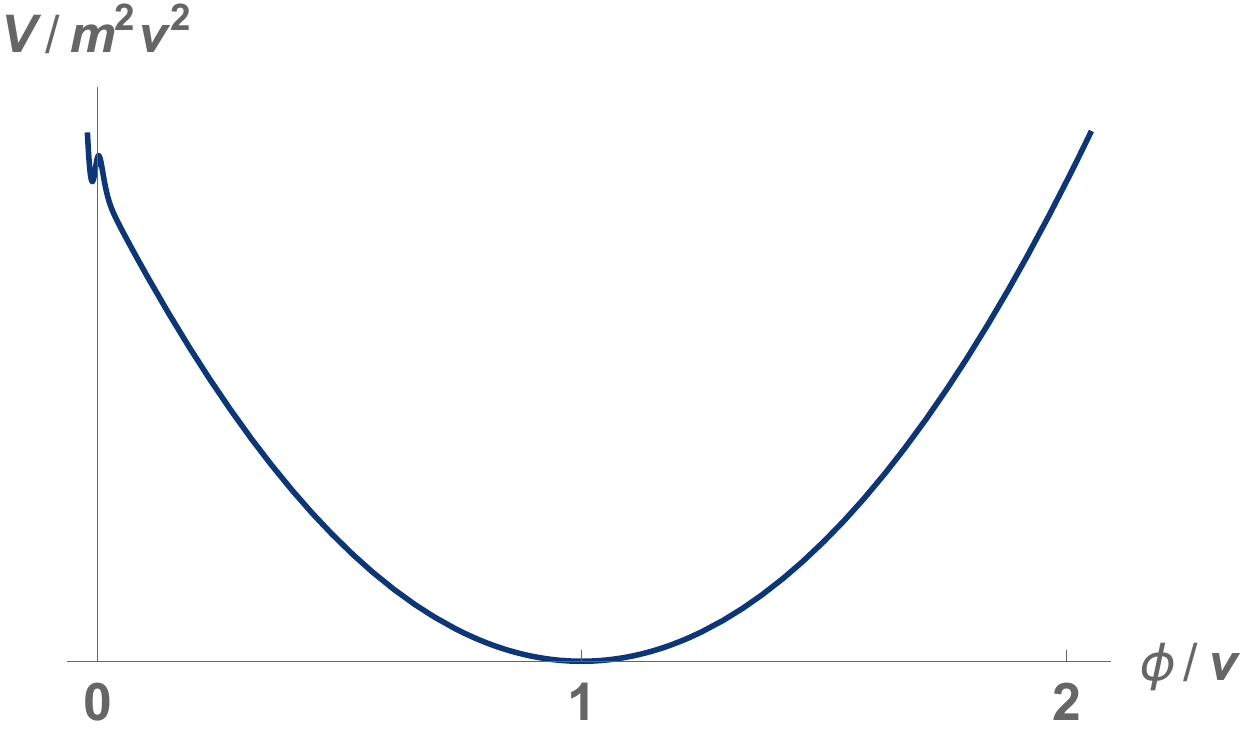}
\caption{\small
Potential shape of the hierarchical model.
}
\label{fig:Hierarchy_V}
\end{center}
\end{figure}

\subsection{Toy model 4: Simple quartic potential}
\label{subsec:Quartic}

\begin{figure}
\begin{center}
\includegraphics[width=0.5\columnwidth]{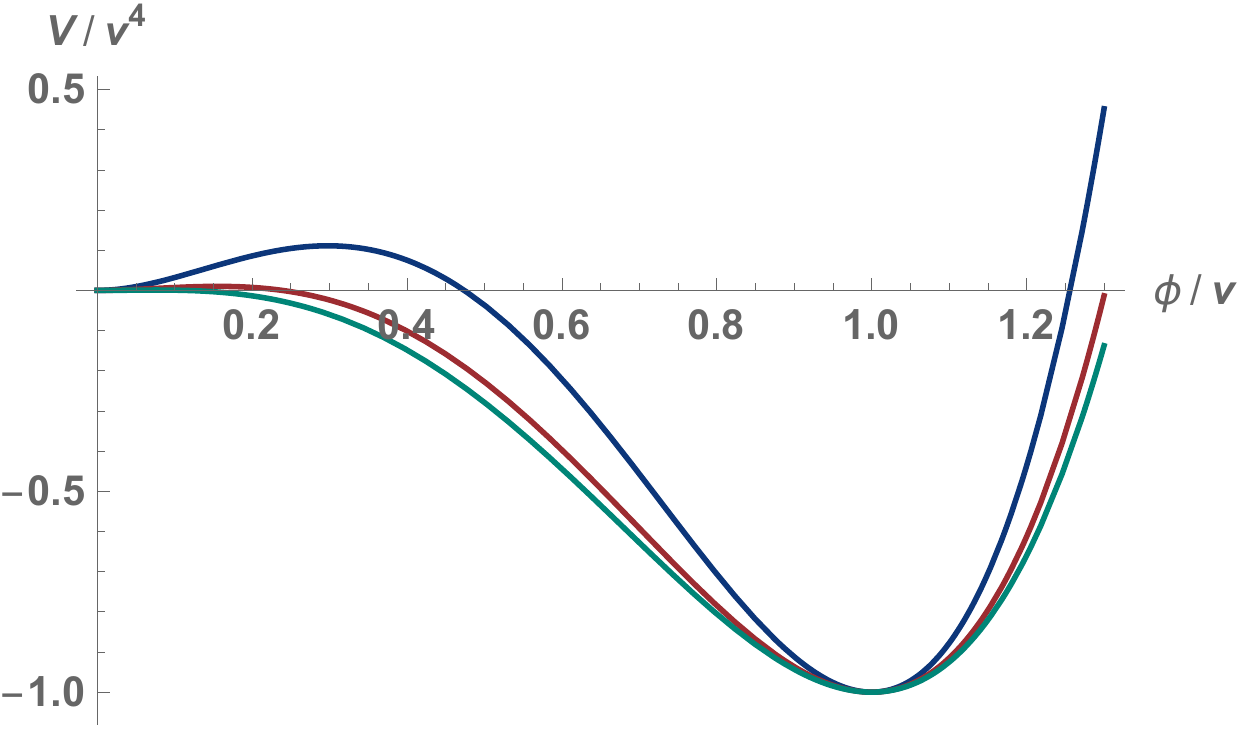}
\caption{\small
Quartic potential for $\epsilon = 0.1$ (blue), $0.01$ (red), and $0.001$ (green).
}
\label{fig:Quartic_V}
\end{center}
\end{figure}

Next, we consider a more realistic potential, namely
\be
V
= 
a v^2 \phi^2 - (2a + 4) \, v \, \phi^3 + (a + 3) \phi^4,
\ee
where the coefficients are chosen so that $V(v) = -v^4$ becomes a local minimum.
This potential takes a local maximum at $\phi / v = a/(2a + 6)$.
We also define the degeneracy parameter $\epsilon$ as
\be
\epsilon
= 
\frac{{\rm (barrier~height)} - {\rm (false~vacuum~height)}}{{\rm (barrier~height)} - {\rm (true~vacuum~height)}}
= 
\frac{a^3(a+4)}{a^3(a+4)+16(a+3)^3}.
\ee
The smaller $\epsilon$ is, the smaller the false vacuum trapping becomes.
In Fig.~\ref{fig:Quartic_V} we plot the potential for $\epsilon = 0.1$, $0.01$, and $0.001$.
The trapping equation (\ref{eq:trapping}) predicts that $\phi$ is trapped at the false vacuum for 
$\epsilon \gtrsim \epsilon_{\rm th} \simeq 0.214$.

For the numerical simulation we use the same definition for the $\gamma$ factor as in Fig.~\ref{fig:gamma_def},
and identify $t_{\rm coll}$ to be $t_\gamma$ for a given value of $\gamma$.
We use $50\gamma \times v x_{\rm coll}$ or $25\gamma \times v x_{\rm coll}$ (both $\propto \gamma^2$) 
points for the spatial discretization for $\gamma \leq 30$ or $\gamma > 30$, respectively.

\begin{figure}
\begin{center}
\includegraphics[width=0.45\columnwidth]{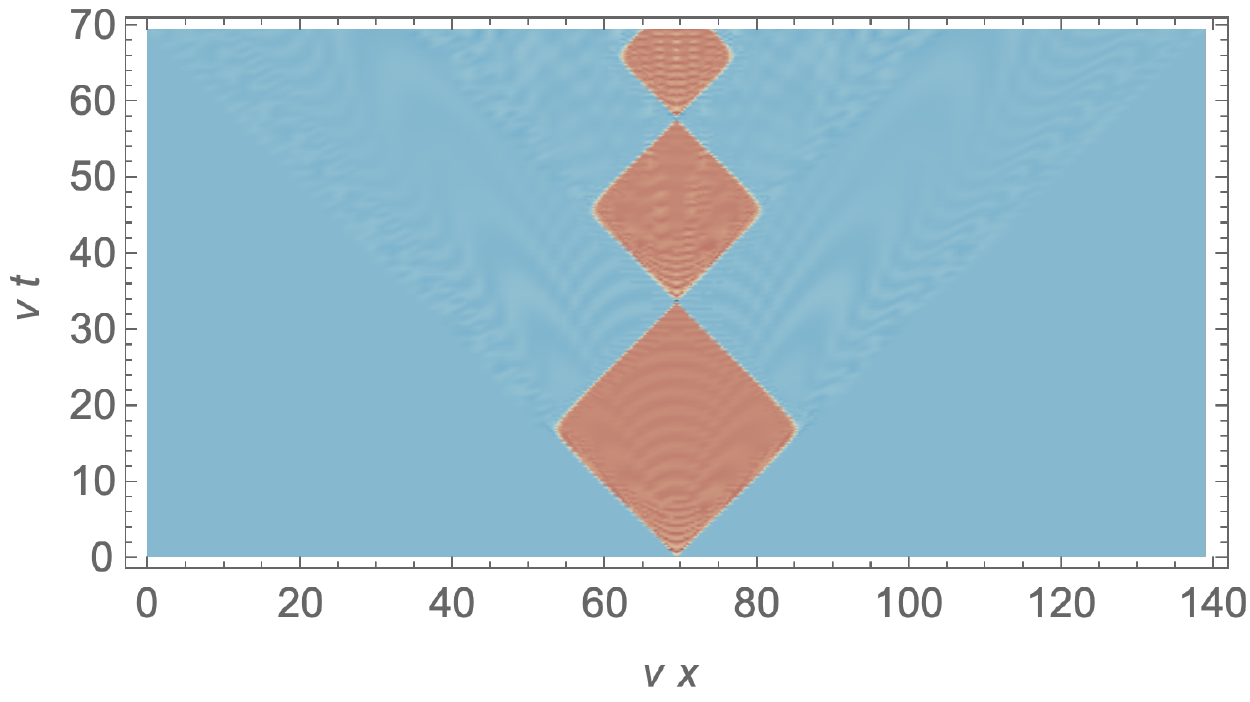}
\hskip 0.5cm
\includegraphics[width=0.4\columnwidth]{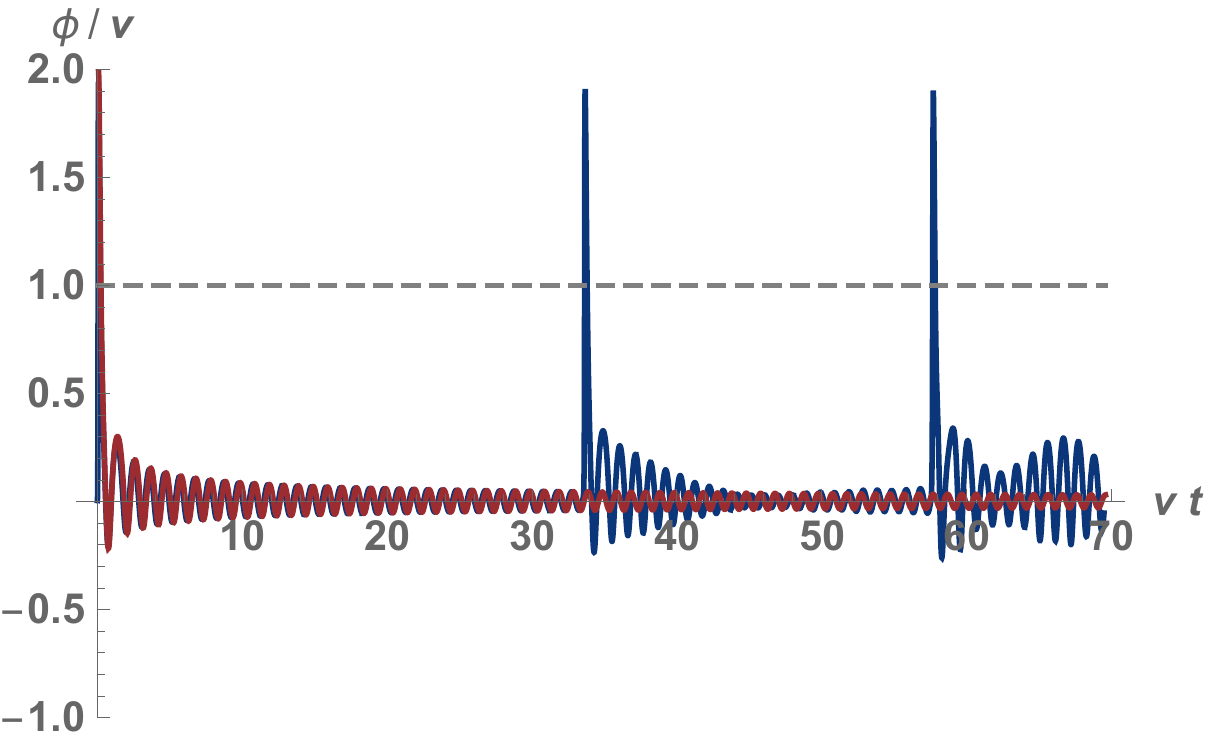}
\vskip 0.3cm
\includegraphics[width=0.45\columnwidth]{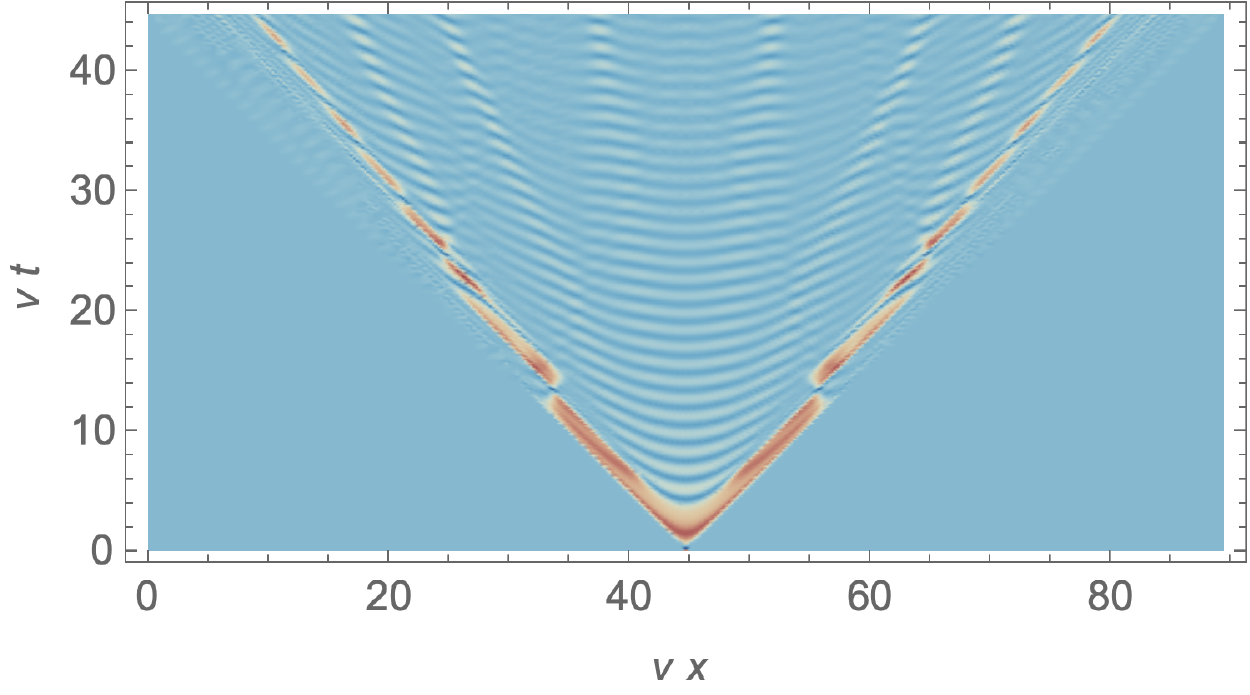}
\hskip 0.5cm
\includegraphics[width=0.4\columnwidth]{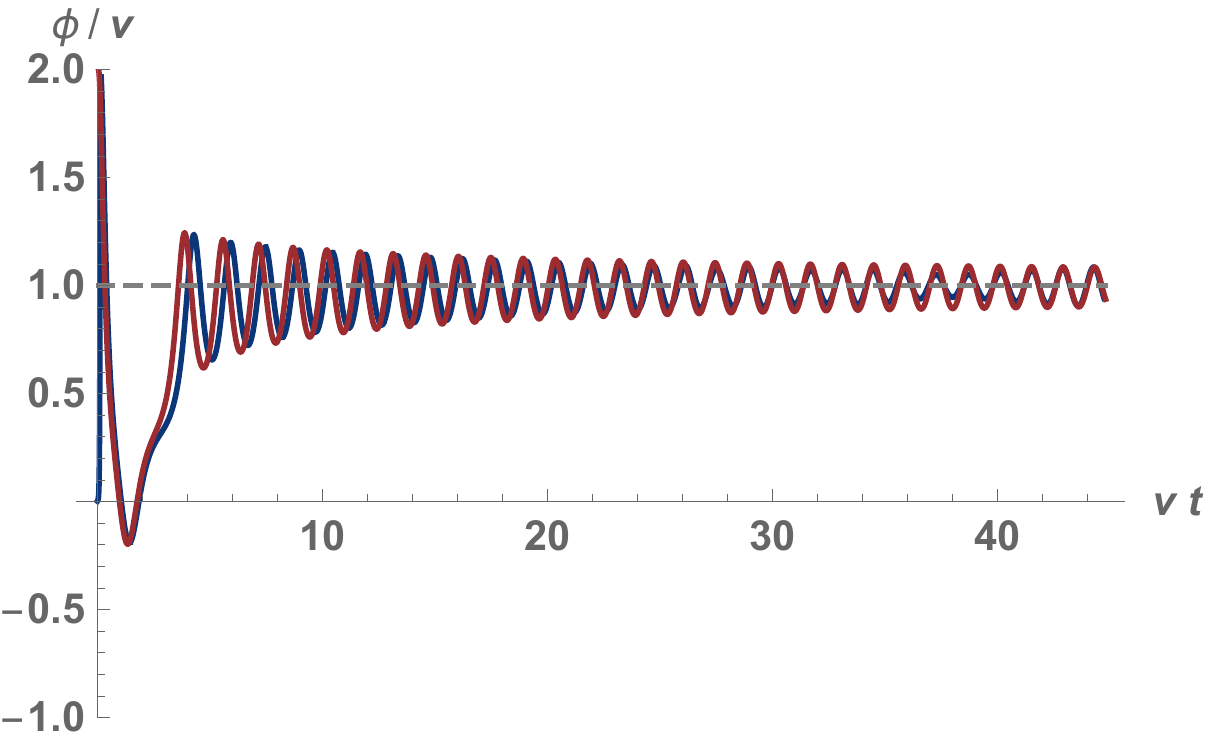}
\vskip 0.3cm
\includegraphics[width=0.45\columnwidth]{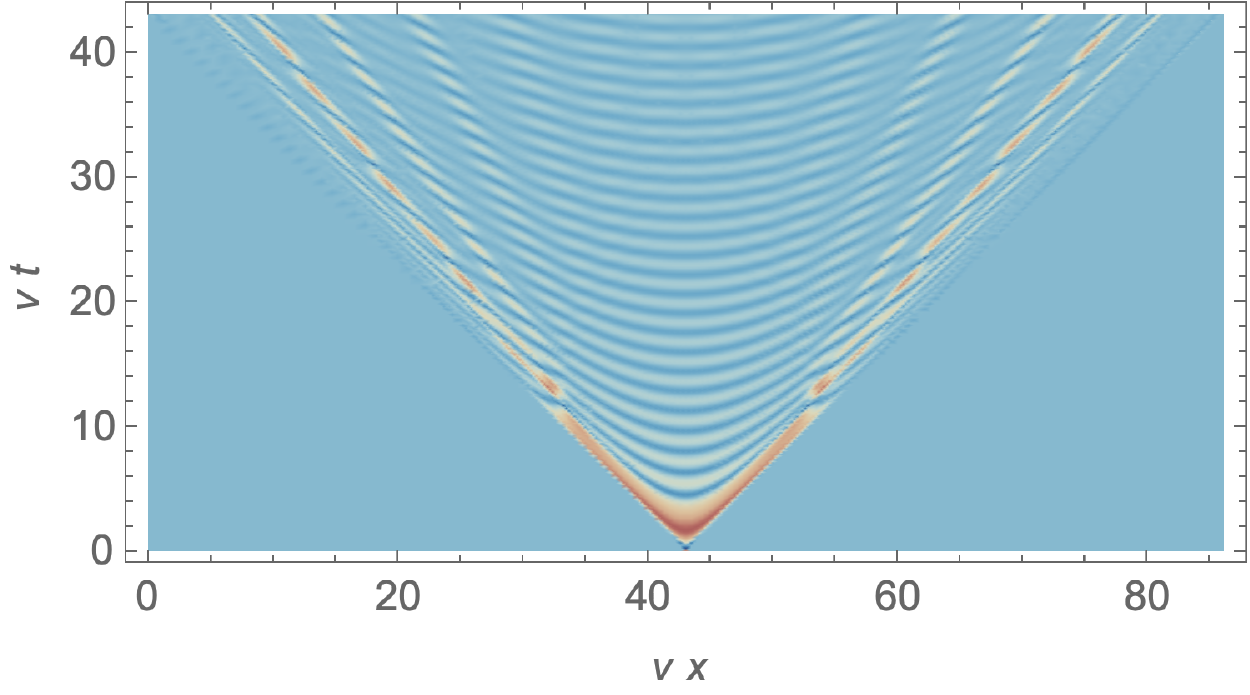}
\hskip 0.5cm
\includegraphics[width=0.4\columnwidth]{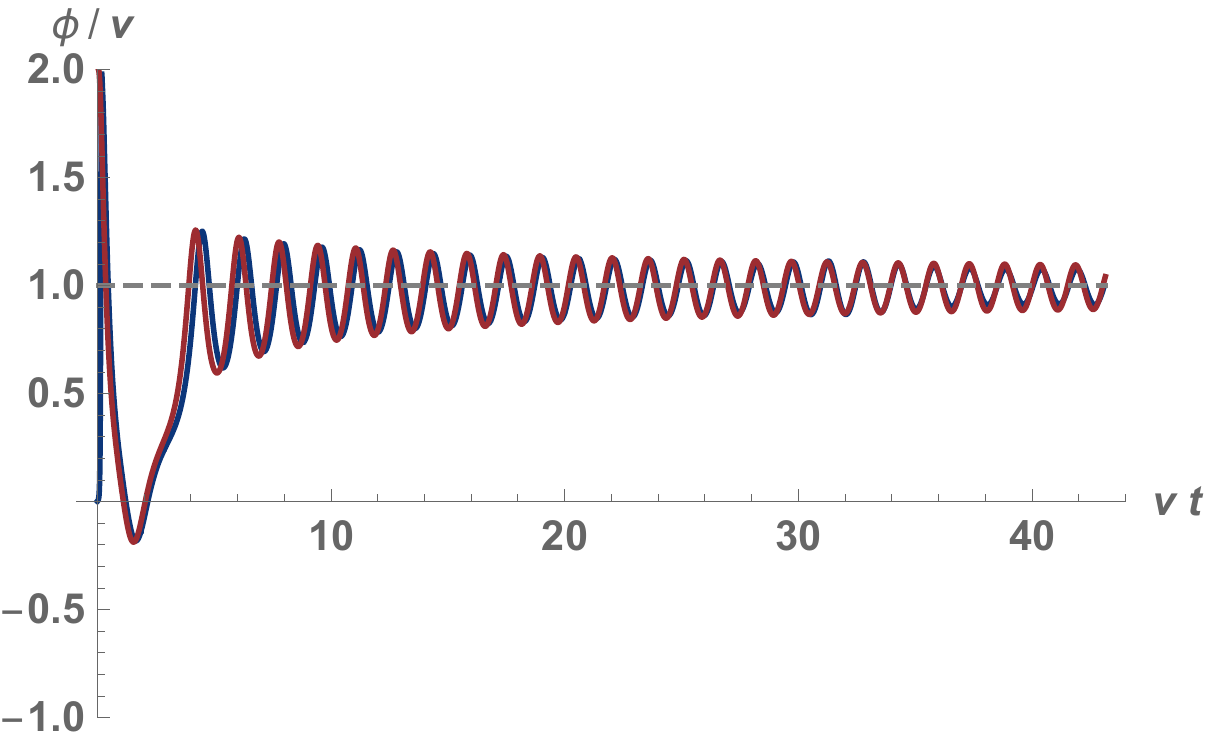}
\caption{\small
Simple quartic potential.
{\bf Left side:} 
Density plot of $\phi$ for $\epsilon = 0.5$, $0.1$, and $0.05$ and $\gamma = 40$ from top to bottom.
Note that false-vacuum trapping is predicted from Eq.~(\ref{eq:trapping}) for $\epsilon = 0.5$.
{\bf Right side:} 
Time evolution of $\phi (t, x = x_{\rm coll})$ for the parameter choice in the left panels.
}
\label{fig:Quartic_phi}
\end{center}
\end{figure}

In Fig.~\ref{fig:Quartic_phi} we plot the time evolution of $\phi$ (left panels) and $\phi (t, x = x_{\rm coll})$ (right panels)
for $\gamma = 40$ and $\epsilon = 0.5$, $0.1$, and $0.05$ from top to bottom.
The blue (red) regions in the left panels correspond to the true (false) vacua, 
while the blue (red) lines in the right panels are the actual (predicted) time evolution.
As predicted by Eq.~(\ref{eq:trapping}),
$\phi$ is trapped at the false vacuum soon after collision for $\epsilon = 0.5$, while it escapes for other values of $\epsilon$.
Also, the diamond-like pattern in the top-left panel can be understood as a consequence of trapping:
Once trapping occurs, the wall receives negative pressure due to the phase difference across it.
The pressure eventually stops the wall motion completely and then inverts it.
The position of the turnback can be estimated by equating 
the energy per surface area at the collision time ($x_{\rm coll} \cdot v^4 / 3$)
with the work per surface area exerted on the wall from collision to turnback ($\Delta x_{\rm turnback} \cdot v^4$)
as $\Delta x_{\rm turnback} \simeq x_{\rm coll}/3$,
which gives a good estimate.
Note that this diamond-like pattern has already been observed in the literature 
(e.g. Refs.~\cite{Kosowsky:1991ua,Konstandin:2011ds,Braden:2014cra,Bond:2015zfa,Cutting:2018tjt}).

\begin{figure}
\begin{center}
\includegraphics[width=0.6\columnwidth]{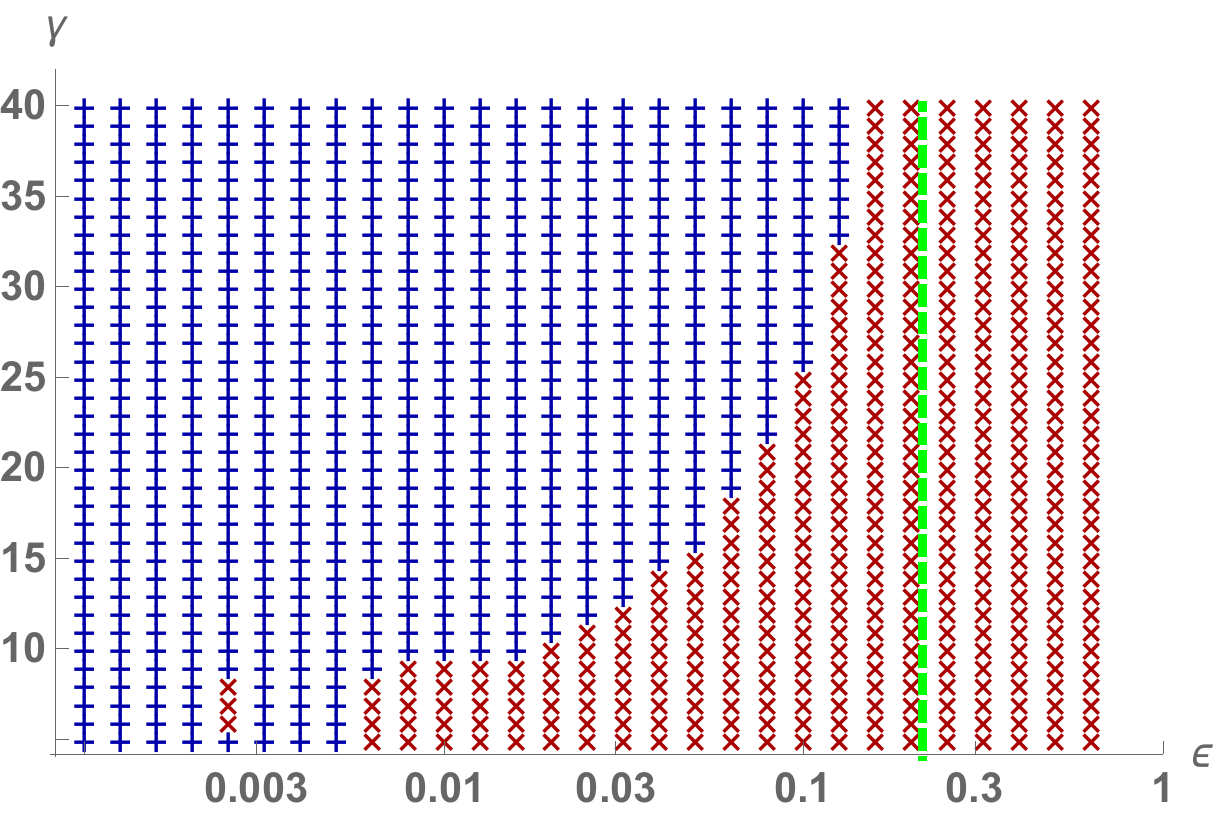}
\caption{\small
Simple quartic potential.
The blue and red points indicate that $\phi$ is trapped at the true and false vacua, respectively.
The green line is the threshold value $\epsilon_{\rm th} \simeq 0.214$ predicted by Eq.~(\ref{eq:trapping}).
}
\label{fig:Quartic_epsilon_gamma}
\end{center}
\end{figure}

Fig.~\ref{fig:Quartic_epsilon_gamma} is the result of our parameter scan.
The blue (red) points are the parameter values where $\phi$ escapes from (is trapped at) the false vacuum\footnote{
The criterion for trapping is as follows.
The 'energy' at the collision point $\left[ (\partial_t {\phi})^2/2 + V(\phi) \right]_{x = x_{\rm coll}}$ decreases after collision.
We numerically calculate the time when it drops to the value of the barrier height
$\left[ (\partial_t {\phi})^2/2 + V(\phi) \right]_{x = x_{\rm coll}} = V (\phi / v = a / (2a + 6))$,
and see whether $\phi$ is in the false or true vacuum side.
We use the same criterion for the quartic $Z_2$ potential as well.
}.
The prediction of the trapping equation $\epsilon \gtrsim \epsilon_{\rm th} \simeq 0.214$ is indicated by the green line.
We see that the boundary between the blue and red points approaches the green line in the large $\gamma$ limit.

\subsection{Toy model 5: Quartic $Z_2$ potential\label{sec:QuarticZ2}}

\begin{figure}
\begin{center}
\includegraphics[width=0.5\columnwidth]{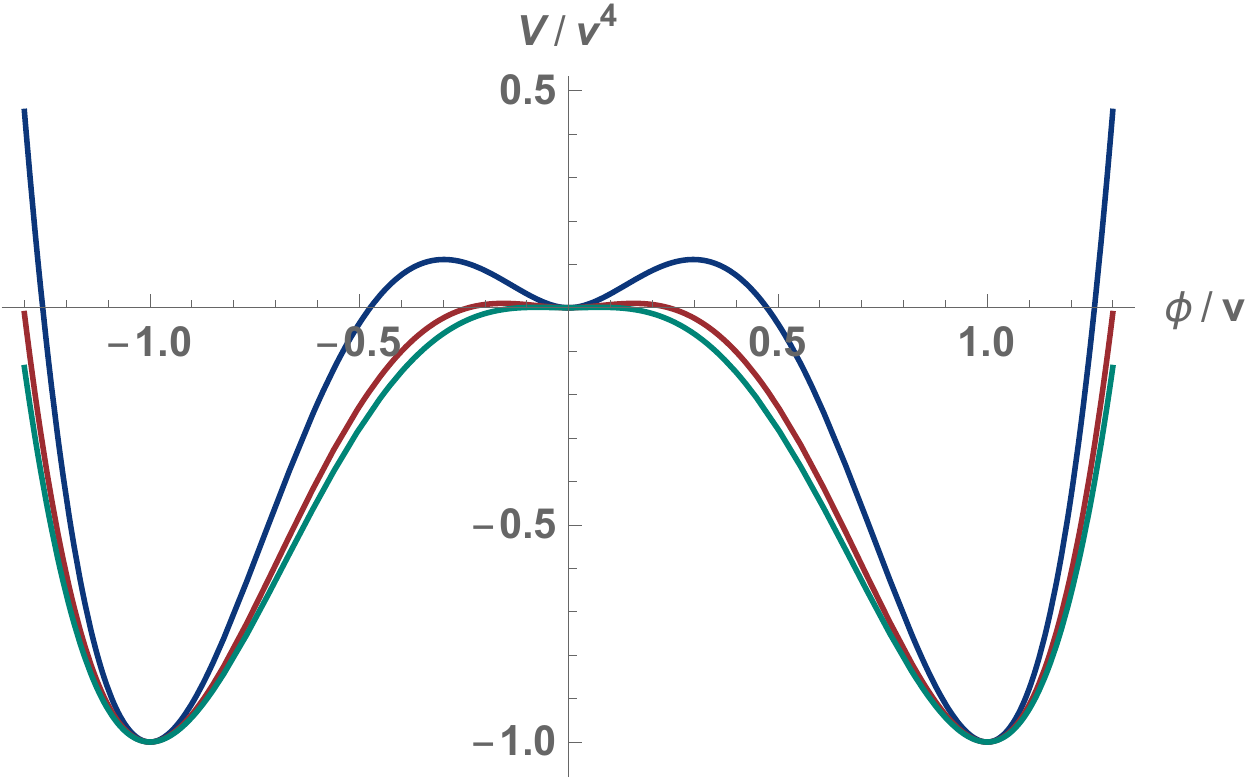}
\caption{\small
Quartic $Z_2$ potential $V$ for $\epsilon = 0.1$ (blue), $0.01$ (red), and $0.001$ (green).
}
\label{fig:QuarticZ2_V}
\end{center}
\vskip 0.7cm
\begin{center}
\includegraphics[width=0.6\columnwidth]{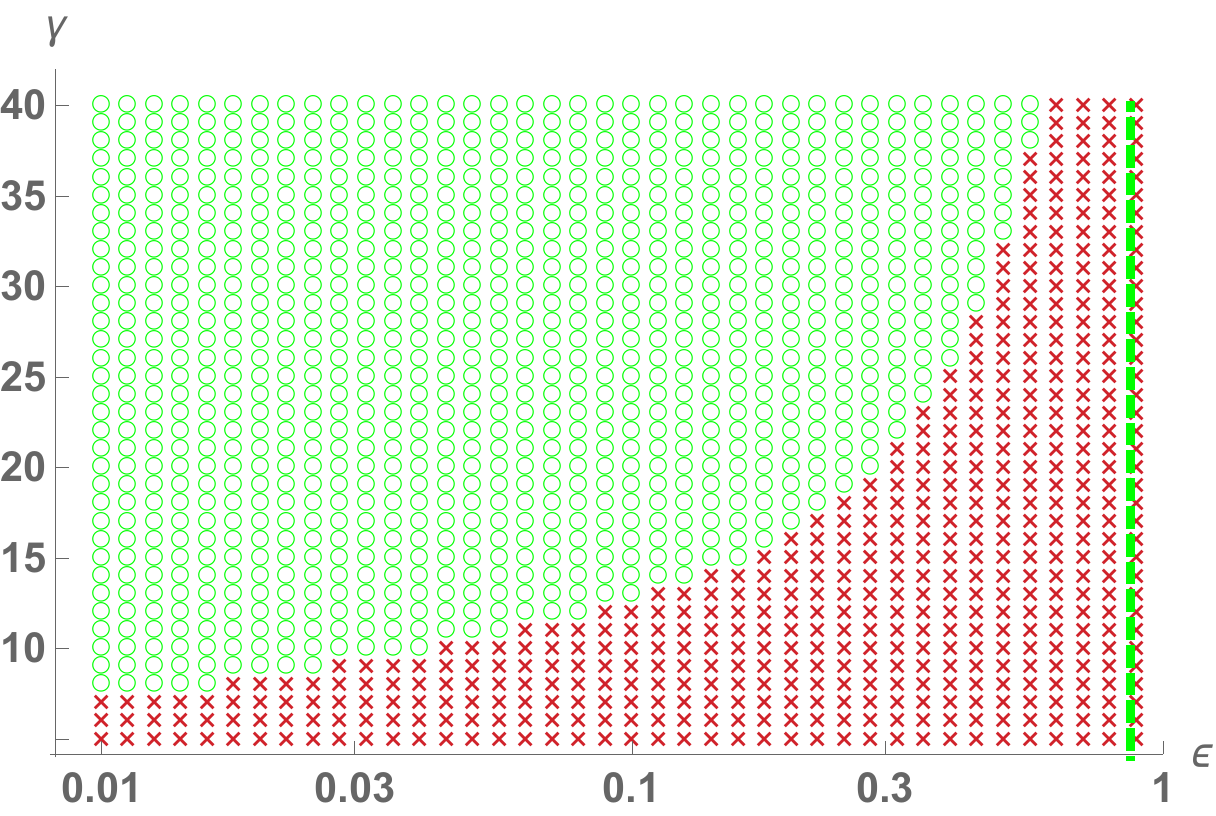}
\caption{\small
Quartic $Z_2$ potential.
The red and green points indicate that $\phi$ is trapped at the symmetric and negative vacua, respectively.
The green line is the threshold value $\epsilon_{\rm th} \simeq 0.867$ predicted by Eq.~(\ref{eq:trapping}).
}
\label{fig:QuarticZ2_epsilon_gamma}
\end{center}
\end{figure}

Finally, we consider a potential similar to the previous one but modified to have $Z_2$ symmetry:
\be
V
= 
a v^2 \, |\phi|^2 - (2a + 4) \, v \,|\phi|^3 + (a + 3)|\phi|^4.
\ee
The potential is plotted in Fig.~\ref{fig:QuarticZ2_V}.
The degeneracy parameter $\epsilon$ is defined in the same way as before.
This setup is not realistic in that a domain wall forms after different-sign configurations collide with each other,
so we study it as a toy model.
In the following we make two same-sign positive configurations collide from opposite directions.
The prediction from Eq.~(\ref{eq:trapping}) is that
$\phi$ is trapped at the negative (opposite) vacuum for $\epsilon < \epsilon_{\rm th}$ with $\epsilon_{\rm th} \simeq 0.867$,
while it settles down to the symmetric one (vanishing VEV) for $\epsilon > \epsilon_{\rm th}$.
There are no parameter values where $\phi$ settles down to the positive vacuum.
This means that the scalar field is likely to be trapped at the opposite vacuum unless the vacua are almost degenerate.
Fig.~\ref{fig:QuarticZ2_epsilon_gamma} is the result of numerical simulation.
The red and green markers mean that $\phi$ is trapped in the zero and negative vacua, respectively,
while the green line is the prediction of the trapping equation (\ref{eq:trapping}).
As indicated from this equation, the scalar field is never trapped at the positive vacuum in the relativistic limit.
Also, the boundary between the red and green regions approaches the green line in the relativistic limit.

\section{Applications of the trapping equation}
\label{sec:appl}
\setcounter{equation}{0}

In the last section we compared the results of the trapping equation with results from $1 + 1$ dimensional simulations. 
We could firmly establish that in the limit of highly-relativistic wall velocities, 
the trapping equation predicts the correct behavior of the scalar field not only qualitatively but also quantitatively quite well. 

\begin{figure}
\begin{center}
\includegraphics[width=0.4\columnwidth]{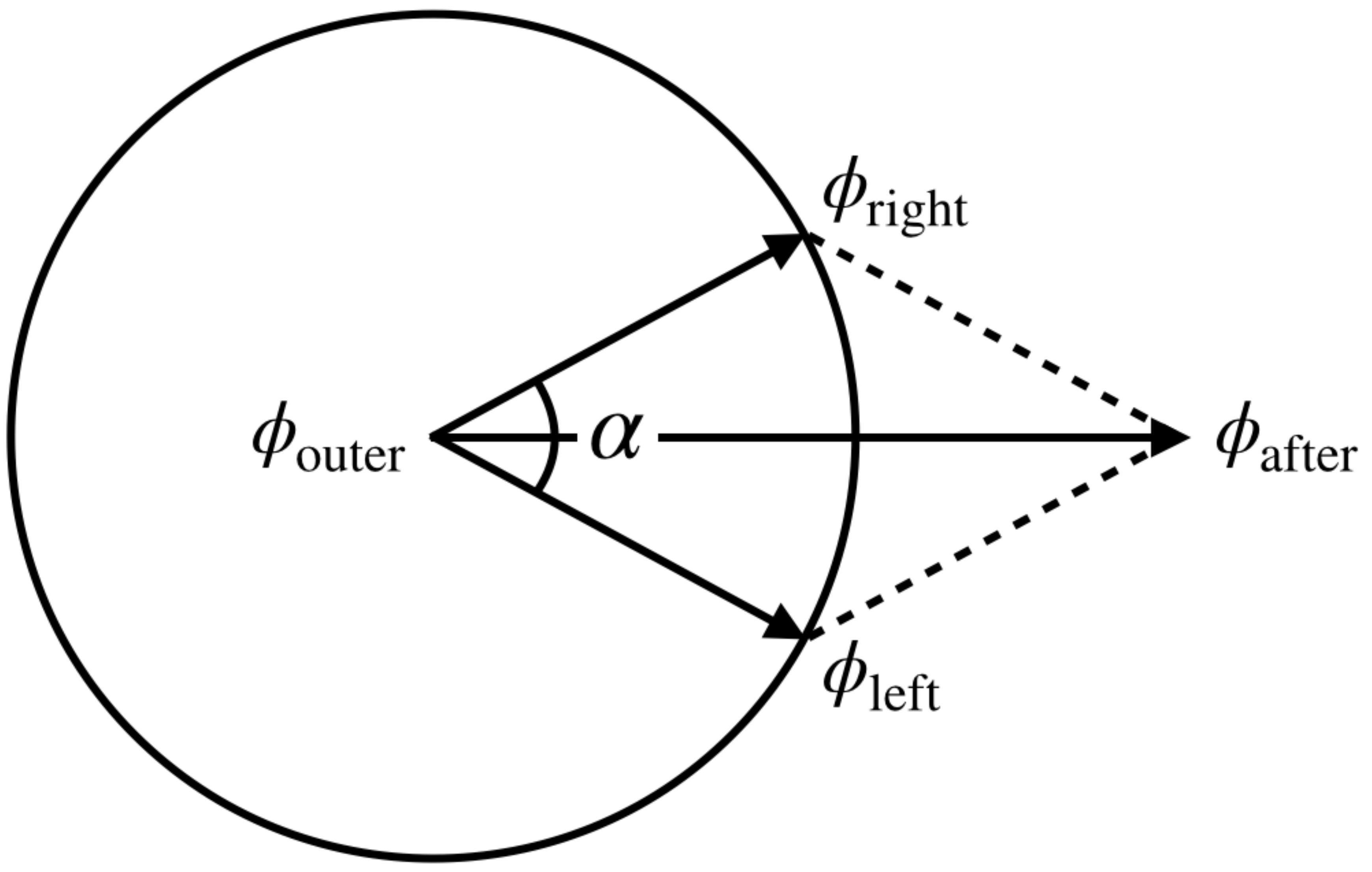}
\caption{\small
Opening angle $\alpha$ for the $U(1)$ case. The $SU(2)$ case is analogous.
}
\label{fig:alpha}
\end{center}
\vskip 0.5cm
\begin{center}
\includegraphics[width=0.6\columnwidth]{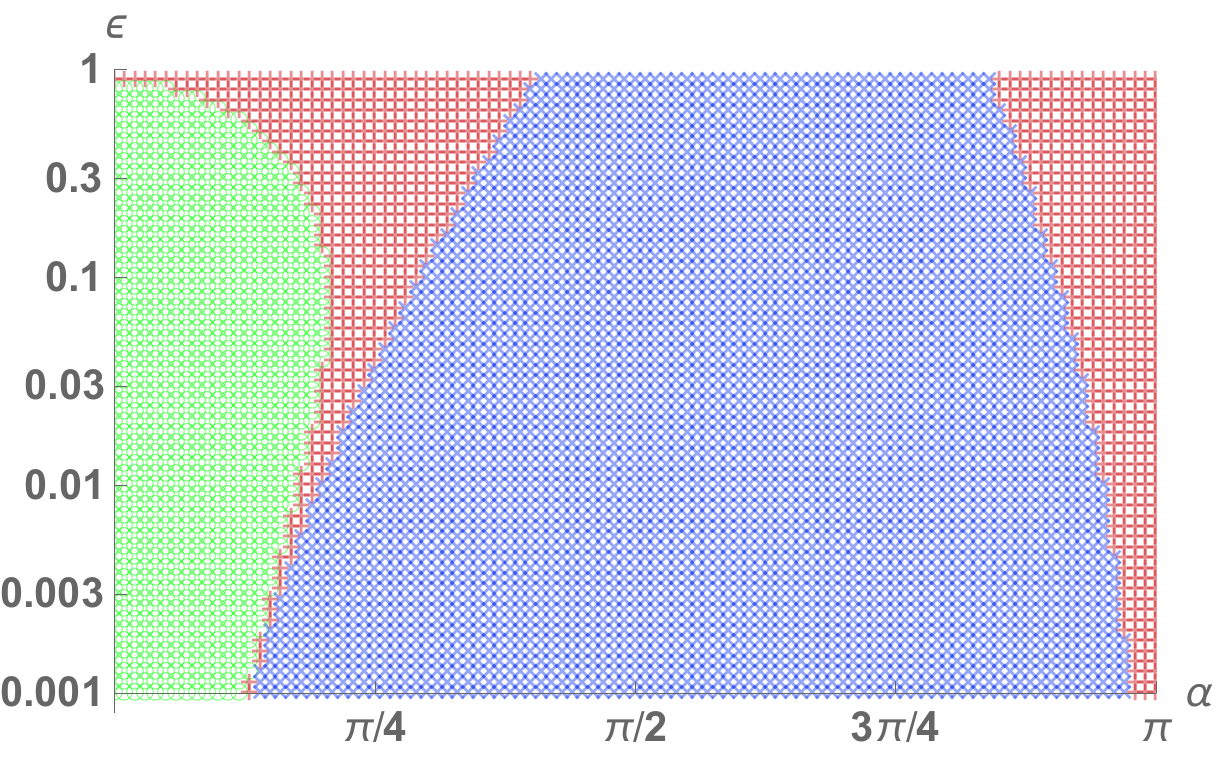}
\caption{\small
Prediction of the trapping equation (\ref{eq:trapping}) with the initial condition (\ref{eq:initial_appl}) 
for $U(1)$ or $SU(2)$ breaking potentials.
The blue, red, and green regions correspond to $\phi = +1$, $0$, and $-1$ for $s \to \infty$.
}
\label{fig:U1}
\end{center}
\end{figure}

In this section we will use the trapping equation to study more complicated setups, 
in particular setups with several scalar fields. 
In this case, lattice simulations might still be possible but solving the trapping equation is almost trivial. 
We discuss trapping for scalar fields with a $U(1)$ and $SU(2)$ global symmetry.
As scalar potential, we use the quartic $Z_2$ potential in Sec.~\ref{sec:test}:
\be
V(|\phi|)
= 
a v^2 \, |\phi|^2 - (2a + 4) \, v \,|\phi|^3 + (a + 3)|\phi|^4.
\ee
In both cases the collision of two solitons is parametrized by the opening angle $\alpha$ between the two configurations.
Correspondingly, the initial condition for the trapping equation (\ref{eq:trapping}) is modified to 
\be
|\phi_{\rm after} - \phi_{\rm outer}|
= 
|\phi_{\rm left} + \phi_{\rm right} - 2\phi_{\rm outer}|
=
2 \, v \, \cos \left( \frac{\alpha}{2} \right),
\label{eq:initial_appl}
\ee
where $\phi_{\rm inner}$ and $\phi_{\rm outer}$ are the value of $\phi$ 
in the broken and symmetric phases, respectively (see Fig.~\ref{fig:alpha}). The case $\alpha =0$ then corresponds to the case studied in the last section while $\alpha=\pi$ corresponds to the collision of the two scalar walls with opposite direction in the $U(1)$ or $SU(2)$ symmetry space.
In the following we take $\phi_{\rm after}$ to be real and positive without loss of generality.

Fig.~\ref{fig:U1} displays the results of the trapping equation (\ref{eq:trapping}) with the initial condition (\ref{eq:initial_appl}).
The blue, red, and green regions denote the regions where $\phi$ is trapped at 
the positive, zero, and negative (opposite) vacua, respectively.
As the false vacuum trapping becomes weaker ($\epsilon \to 0$),
the scalar field becomes less likely to be trapped at the false vacuum.

\section{Energy dynamics after collisions}
\label{sec:energy}
\setcounter{equation}{0}

Finally we discuss the energy dynamics after the collision.
This is important because the energy distribution determines observable signatures like the GW spectrum.
Indeed, the GW spectrum takes quite different forms
when the scalar field instantly lose energy at the collision point~\cite{Kosowsky:1992vn,Huber:2008hg,Jinno:2016vai}
and when energy propagates even after collisions~\cite{Jinno:2017fby,Konstandin:2017sat}.
Therefore, our main interest lies in the degree of energy localization.
For later purpose let us first define $d_R$ as follows:
\begin{align}
d_R
&\equiv 
{\rm minimum~value~of~the~spatial~interval}
\nonumber \\
&~~~~
{\rm ~in~which~fraction~}R{\rm ~of~the~total~energy~is~localized}.
\end{align}
For example, $d_{0.5}$ and $d_{0.8}$ respectively mean that 
we can find $d_{0.5}$ and $d_{0.8}$ intervals in which $50\%$ and $80\%$ of the total energy is localized.
In the following we present the ratio $d_R (t = t_{\rm end}) / d_R (t = t_{\rm coll})$,
which parametrizes the degree of wall thickening during evolution from $t_{\rm coll}$ to $t_{\rm end}$.

\begin{figure}
\begin{center}
\includegraphics[width=0.45\columnwidth]{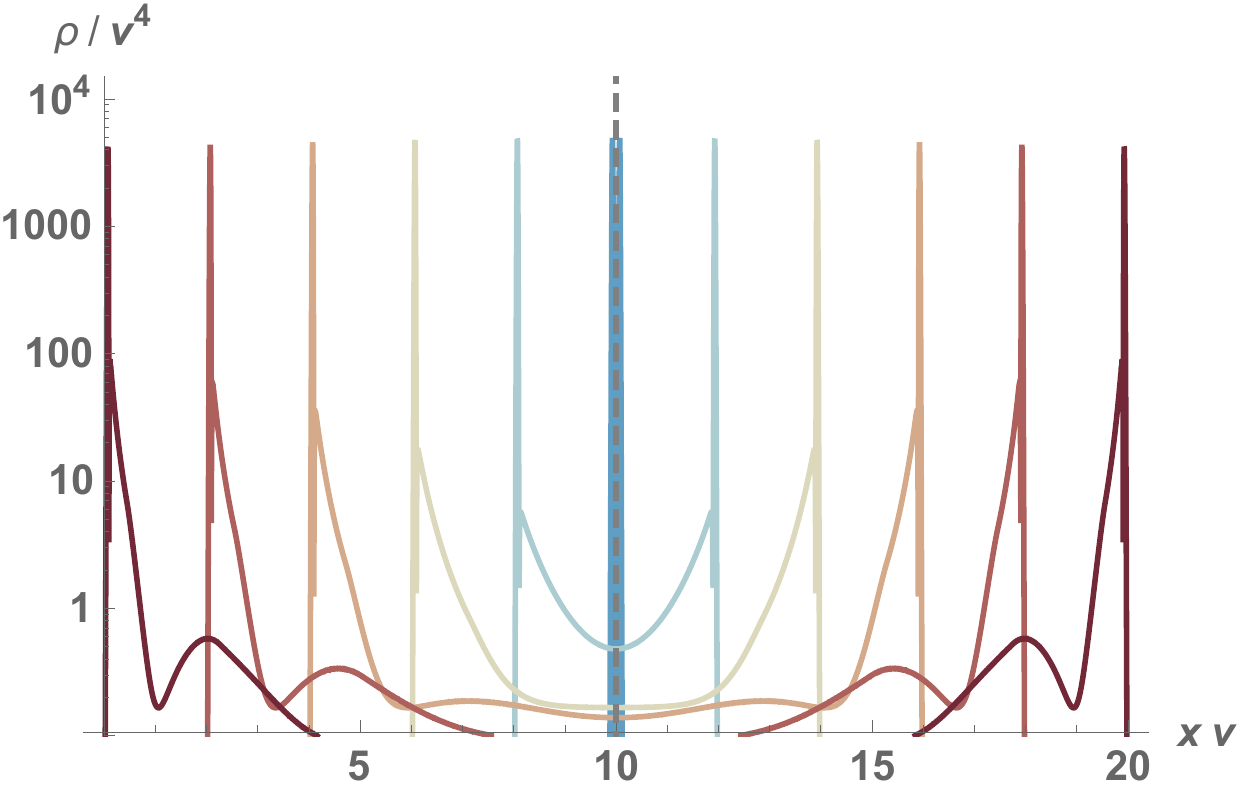}
\includegraphics[width=0.45\columnwidth]{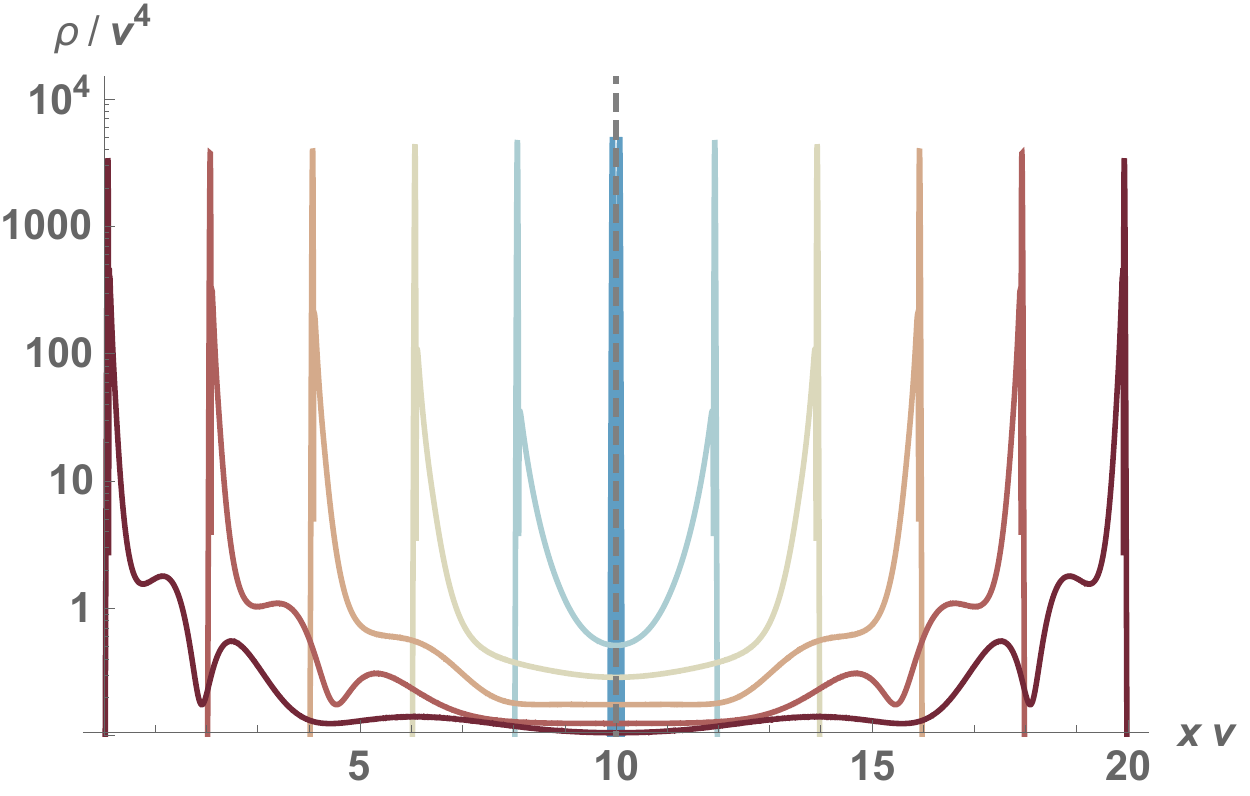}
\caption{\small
Modified $Z_2$: Time evolution of $\rho$ for $\lambda = 0.1$ (left) and $0.3$ (right) with the $\gamma$ factor of $100$.
The system evolves from the blue lines to the red lines.
These plots correspond to the parameter choice of Fig.~\ref{fig:Z2mod_phi}.
The collision occurs at $x \simeq 10/v$, where we impose reflecting boundary conditions.
\label{fig:Z2_rho}
}
\vskip 0.5cm
\includegraphics[width=0.45\columnwidth]{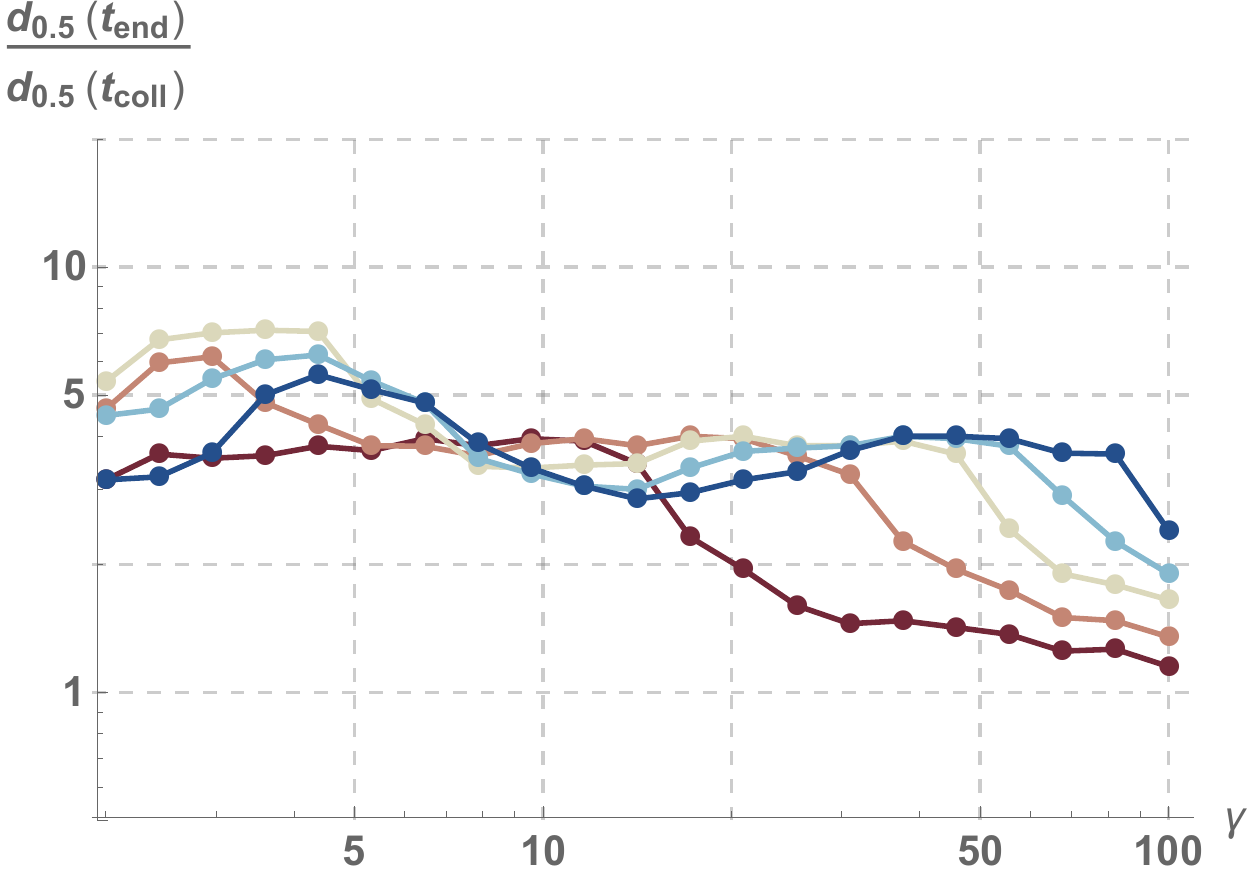}
\includegraphics[width=0.45\columnwidth]{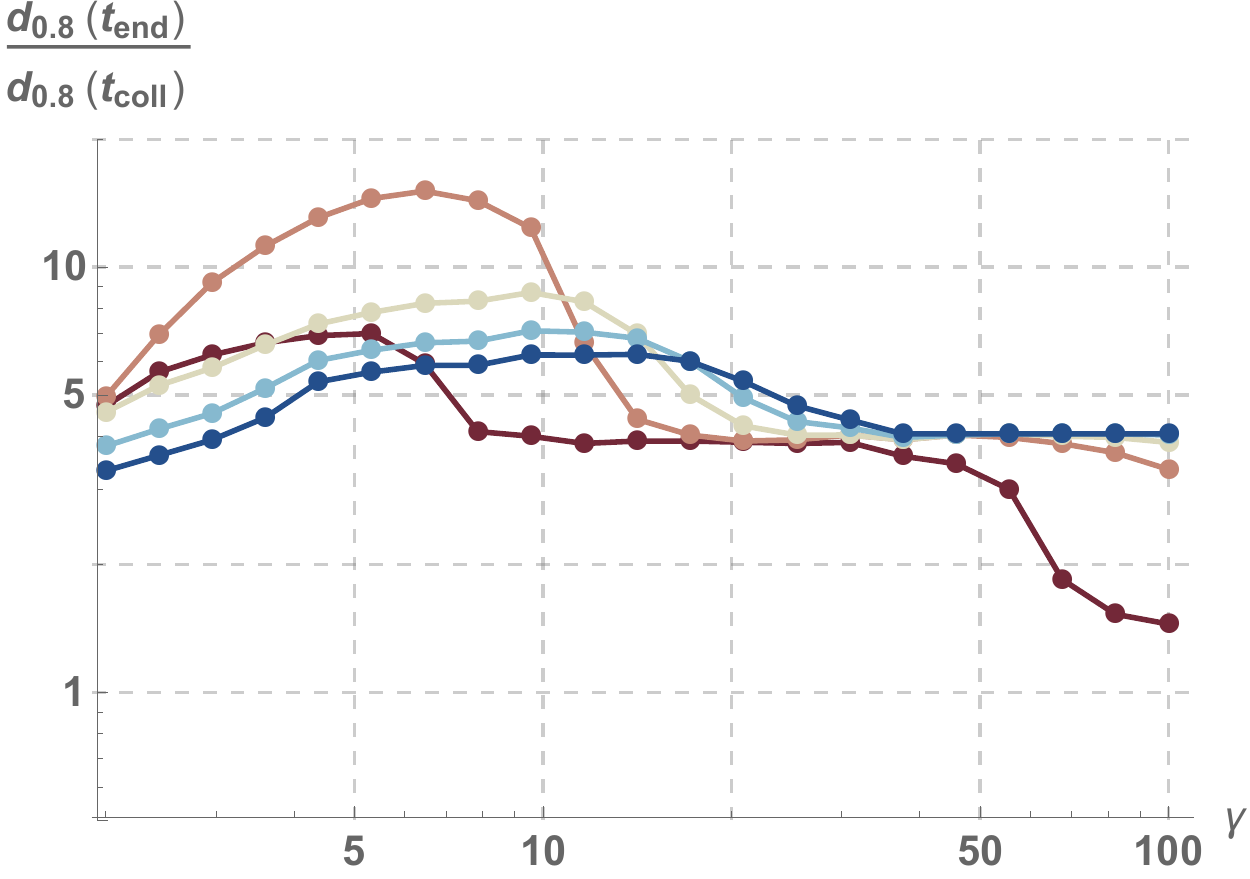}
\caption{\small
Modified $Z_2$: Thickness ratio 
$d_{0.5} (t_{\rm end}) / d_{0.5} (t_{\rm coll})$ (left) and $d_{0.8} (t_{\rm end}) / d_{0.8} (t_{\rm coll})$ (right)
for $\lambda = 0.1$, $0.2$, $0.3$, $0.4$, and $0.5$ from the blue to the red lines.
\label{fig:Z2mod_ThicknessRatio}
}
\end{center}
\end{figure}

{\bf Modified $Z_2$:}
As a first example, consider the modified $Z_2$ potential in Sec.~\ref{subsec:Z2}. The time evolution of the energy density $\rho$ is displayed in Fig.~\ref{fig:Z2_rho}
for $\lambda = 0.1$ (left) and $0.3$ (right).
The $\gamma$ factor is taken to be $100$.
Reflecting boundary conditions are imposed at $x \simeq 10/v$, and the system evolves from the blue to the red lines.
Note that these parameters are the same as Fig.~\ref{fig:Z2mod_phi}.
The left panel corresponds to the case where the scalar field stays at the positive vacuum,
while in the right panel it bounces back to the negative vacuum.
We see that in both cases the energy is localized at the wall front even in the last time slice (the outermost profiles).

In Fig.~\ref{fig:Z2mod_ThicknessRatio} we plot the ratio
$d_{0.5} (t_{\rm end}) / d_{0.5} (t_{\rm coll})$ (left) and $d_{0.8} (t_{\rm end}) / d_{0.8} (t_{\rm coll})$ (right).
For any value of $\lambda$, the thickness ratio keeps ${\mathcal O}(1)$ values or gradually decreases 
as $\gamma$ increases.
Note that the simulation time corresponds to the typical bubble size in a realistic situation.
Since the initial wall thickness decreases as $\gamma^{-1}$,
this means that the wall thickness after propagating over a distance comparable to bubble radius
also decreases as $\gamma^{-1}$, regardless of whether $\phi$ bounces back to the old phase or not.

\begin{figure}
\begin{center}
\includegraphics[width=0.45\columnwidth]{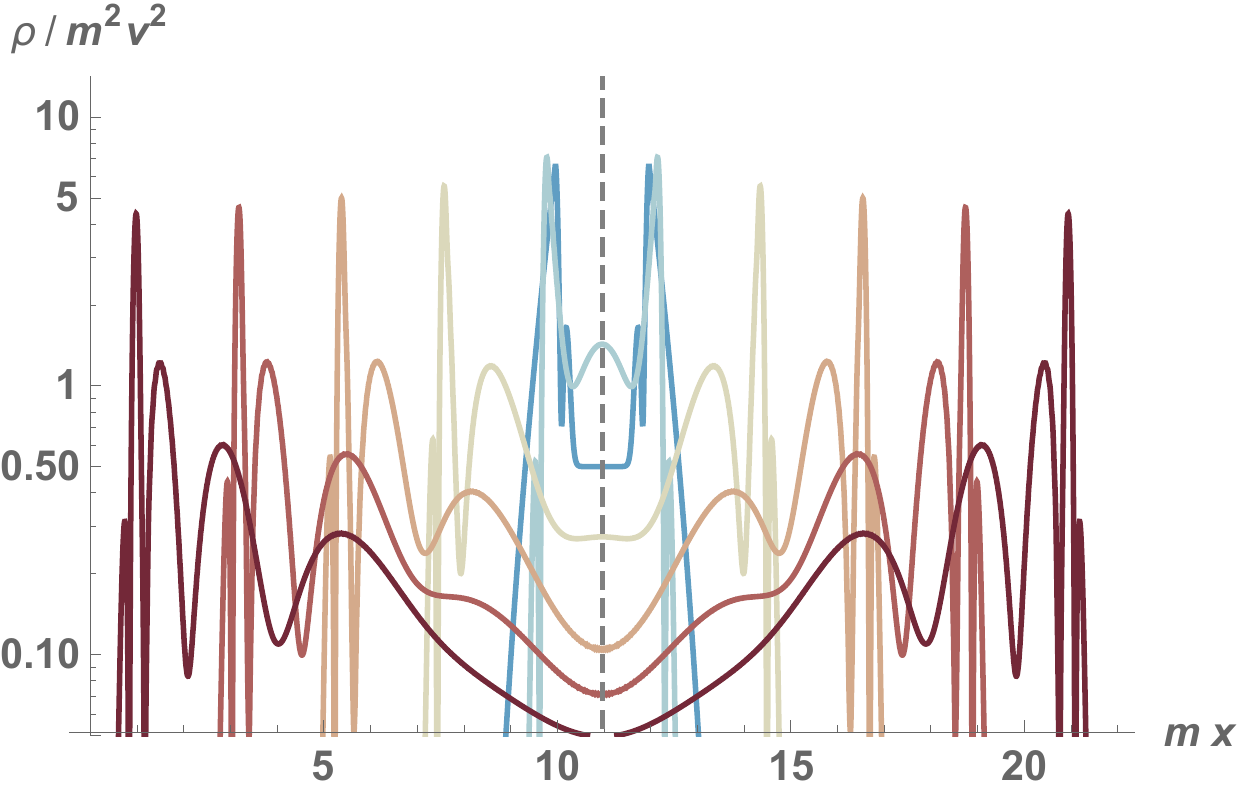}
\includegraphics[width=0.45\columnwidth]{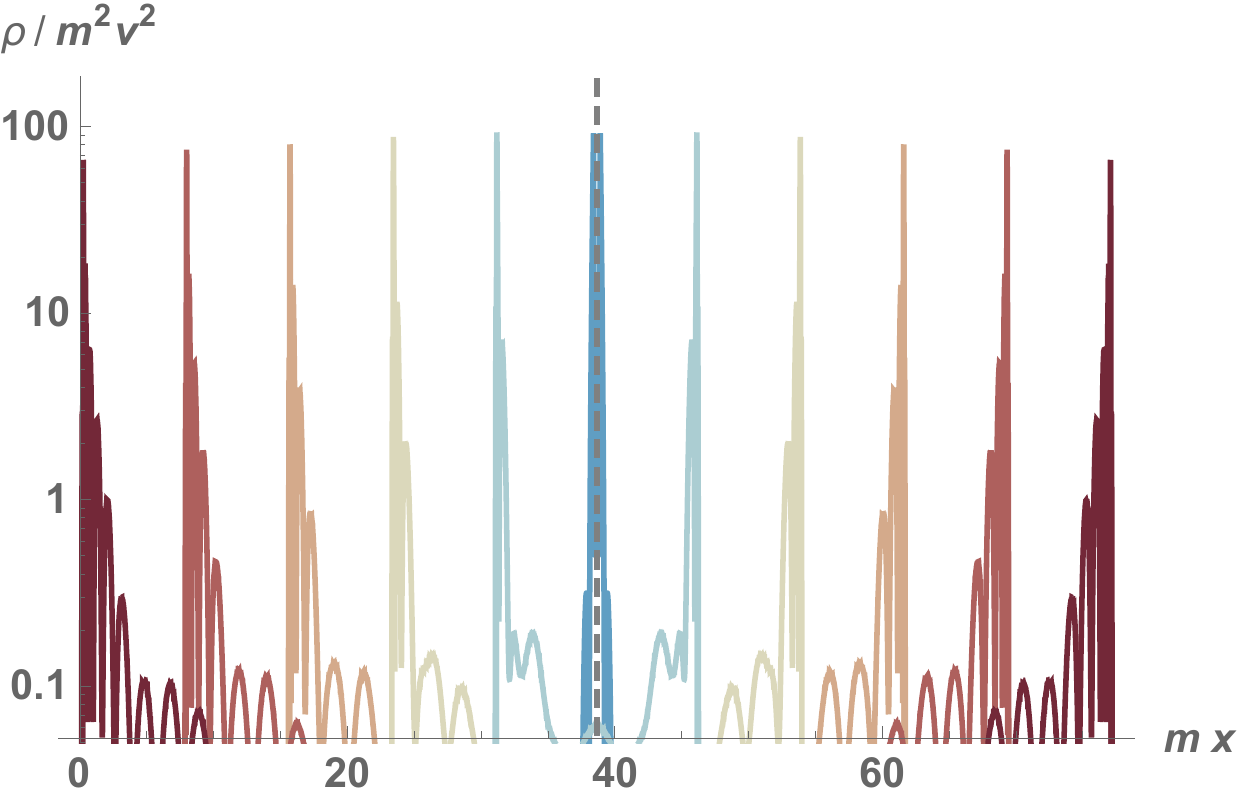}
\caption{\small
Hierarchical potential: 
Time evolution of $\rho$ for $\gamma = 5$ (left) and $20$ (right).
The collision occurs at the center, where we impose reflecting boundary conditions,
and the system evolves from the blue to the red lines.
\label{fig:Hierarchy_rho}
}
\vskip 1cm
\includegraphics[width=0.6\columnwidth]{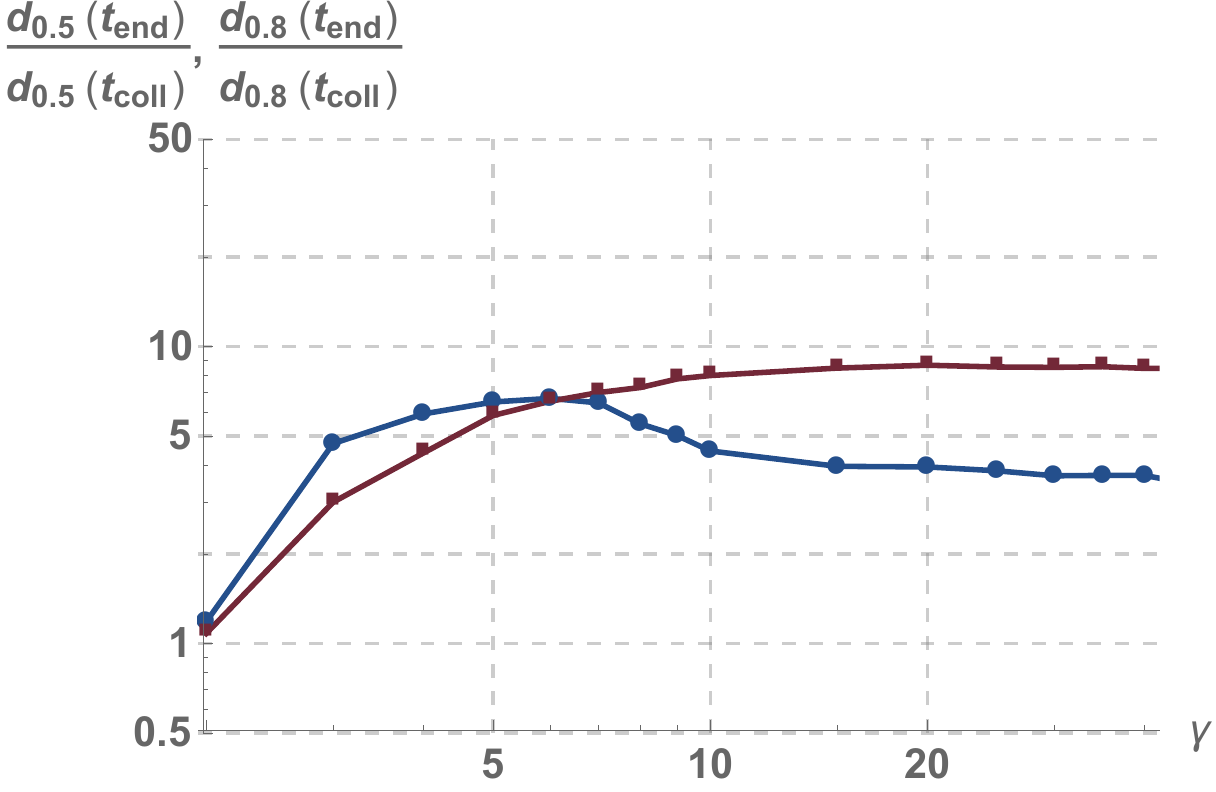}
\caption{\small
Thickness ratio $d_{0.5}(t = t_{\rm end})/d_{0.5}(t = t_{\rm coll})$ (blue)
and $d_{0.8}(t = t_{\rm end})/d_{0.8}(t = t_{\rm coll})$ (red) for the hierarchical potential.
\label{fig:Hierarchy_ThicknessRatio}
}
\end{center}
\end{figure}

{\bf Hierarchical potential:}
Next let us discuss the hierarchical potential from Sec.~\ref{subsec:Quadratic}.
In Fig.~\ref{fig:Hierarchy_rho} we plot the time evolution of the energy density $\rho$ for $\gamma = 5$ (left) and $20$ (right).
Just as in the previous example, the energy localization is still strong even in the last time slice.

In Fig.~\ref{fig:Hierarchy_ThicknessRatio}
we show the thickness ratio $d_{0.5}(t = t_{\rm end})/d_{0.5}(t = t_{\rm coll})$ (blue)
and $d_{0.8}(t = t_{\rm end})/d_{0.8}(t = t_{\rm coll})$ (red).
We see that the thickness ratios approach ${\mathcal O}(1)$ values as $\gamma$ increases.
Again, since the simulation time corresponds to the typical bubble size in a realistic situation,
we expect that the wall thickness remains to be ${\rm (particle~physics~scale)}^{-1}$
even after the scalar field propagates over the typical bubble size.

\begin{figure}
\begin{center}
\includegraphics[width=0.45\columnwidth]{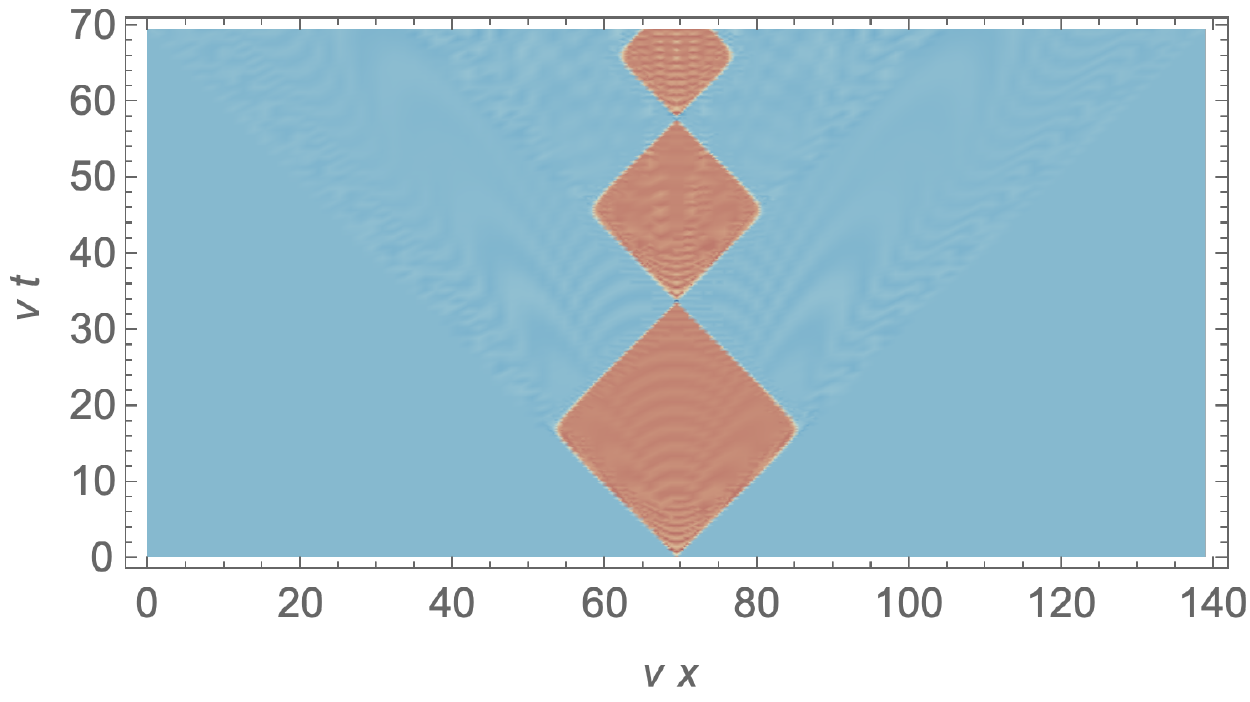}
\includegraphics[width=0.45\columnwidth]{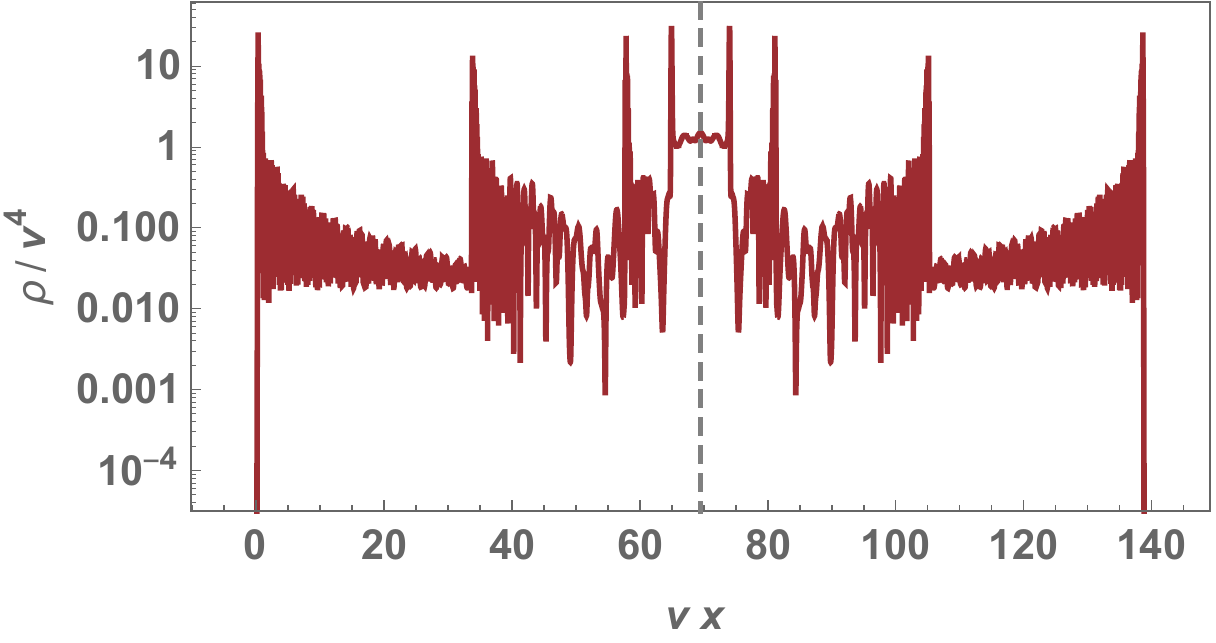}
\caption{\small
Time evolution of $\phi$ (left) and the energy density at the simulation end $\rho(t = t_{\rm end})$ (right)
for the quartic potential with $\epsilon = 0.5$ and $\gamma = 40$.
The collision occurs at the position of the dashed line in the right panel.
The trapping equation (\ref{eq:trapping}) predicts that $\phi$ is trapped at the false vacuum at the initial stage.
}
\label{fig:Quartic_trapped}
\end{center}
\vskip 0.5cm
\begin{center}
\includegraphics[width=0.45\columnwidth]{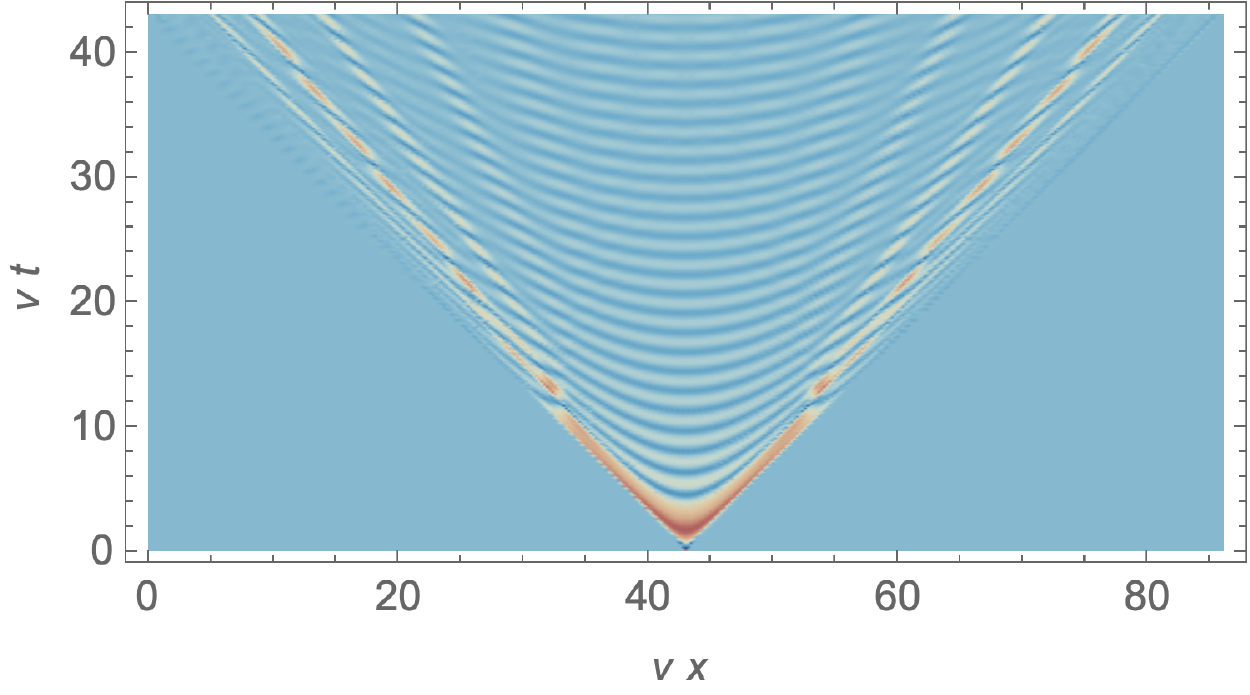}
\includegraphics[width=0.45\columnwidth]{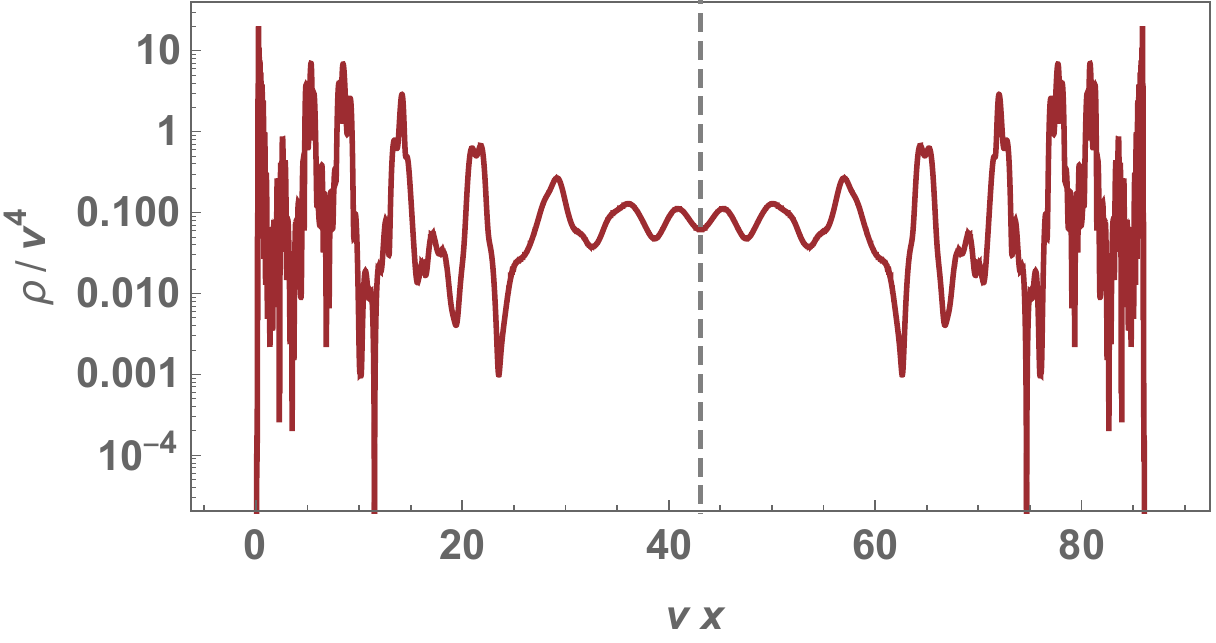}
\caption{\small
The same as Fig.~\ref{fig:Quartic_trapped} except that $\epsilon = 0.05$ and $\gamma = 40$.
The trapping equation (\ref{eq:trapping}) predicts that $\phi$ escapes from the false vacuum.
}
\label{fig:Quartic_escaped}
\end{center}
\vskip 0.5cm
\begin{center}
\includegraphics[width=0.45\columnwidth]{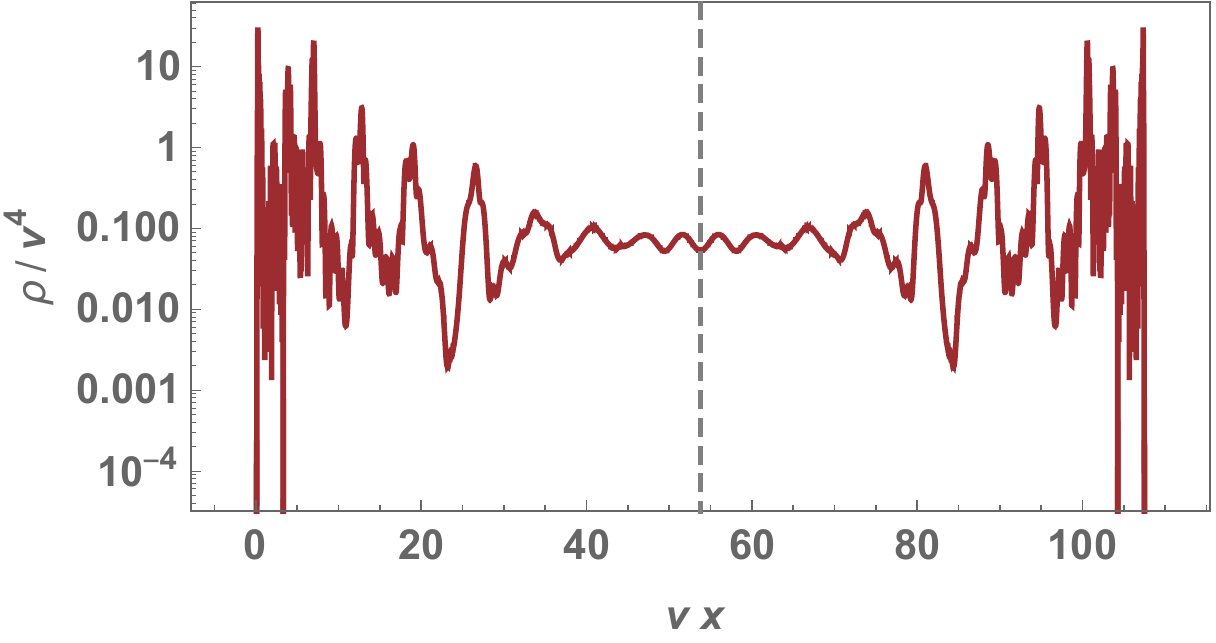}
\includegraphics[width=0.45\columnwidth]{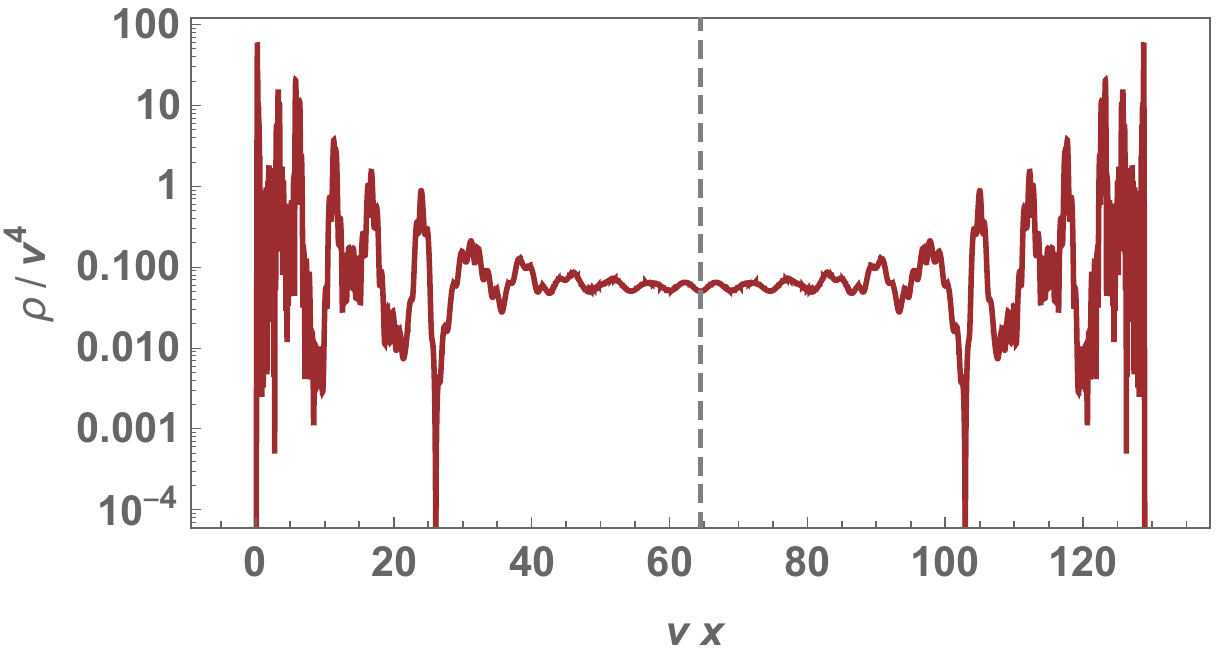}
\caption{\small
How the right panel of Fig.~\ref{fig:Quartic_escaped} changes for different values of $\gamma$.
The value of $\gamma$ is chosen to be $50$ and $60$ for the left and right panels, respectively.
The inner peaks merge with the outermost peaks as $\gamma$ increases.
}
\label{fig:Quartic_escaped_limit}
\end{center}
\end{figure}

{\bf Quartic potential:}
Let us finally study the quartic potential in Sec.~\ref{subsec:Quartic}.
In this case the behavior of the scalar field is much more complicated than the previous two examples.
We take two parameter points $\epsilon = 0.5$ (Fig.~\ref{fig:Quartic_trapped})
and $0.05$ (Figs.~\ref{fig:Quartic_escaped} and \ref{fig:Quartic_escaped_limit}).
The trapping equation (\ref{eq:trapping}) predicts that the scalar field is trapped at (escapes from) 
the false vacuum for the former (latter) potential at the initial stage.

In Fig.~\ref{fig:Quartic_trapped} we show the case with a sizable barrier between the two phases ($\epsilon = 0.5$).
The left panel (the same as the top-right panel of Fig.~\ref{fig:Quartic_phi}) shows the time evolution of $\phi$,
while the right panel is the energy density distribution at $t = t_{\rm end}$.
The $\gamma$ factor is taken to be $40$.
We see from the left panel that the scalar field is indeed trapped in the false vacuum as Eq.~(\ref{eq:trapping}) predicts. 
We also see several collisions at $vt \simeq 0$, $34$, and $58$ caused by the trapping.
Since the scalar field is again trapped in the false vacuum, the large pressure across the wall decelerates 
the wall and then accelerates it again for the subsequent collision.
These multiple collisions result in the energy distribution in the right panel at the simulation end.
The three peaks (from outside to inside) come from the first, second and third collisions, respectively,
while the energy localization at the center is the effect of trapping still continuing at the simulation end.
Interestingly, the outermost peak does not dominate the energy of the system:
it carries only $0.241$ of the total energy, and this fraction does not change significantly even if $\gamma$ increases.
Indeed it takes $0.245$ and $0.246$ for $\gamma = 50$ and $60$, respectively.
In addition, the distance between the outermost and inner peaks is stable against the change in $\gamma$,
since it is determined by the condition ``(energy released until just before collision) $\simeq$ 
(energy stored in the false vacuum trapping)".
Therefore, we conclude that the energy does not localize at the front if trapping occurs\footnote{
However, note two things: (1) The distance between the outermost and inner peaks is 
${\mathcal O}(0.1) \times {\rm (bubble~radius)}$, which will not change significantly even in $3 + 1$ dimensional collisions.
(2) Both peaks propagate at relativistic speeds.
These mean that, after the energy peaks propagate over a distance much longer than the bubble radius at collisions,
their distance is much shorter than the radius of the bubble-like structures.
Therefore, the IR structure pointed out in Ref.~\cite{Jinno:2017fby} may appear in the GW spectrum.
}.

In Fig.~\ref{fig:Quartic_escaped} we show the case with a rather small barrier ($\epsilon = 0.05$).
The $\gamma$ factor is taken to be the same as before, $\gamma = 40$.
We see that the trapping does not occur, as Eq.~(\ref{eq:trapping}) predicts.
In contrast to Fig.~\ref{fig:Quartic_trapped}, the outermost peaks are the highest ones in this case.
The subsequent peaks carry a non-negligible fraction of the total energy, 
but they merge with the outermost ones in the large $\gamma$ limit.
Fig.~\ref{fig:Quartic_escaped_limit} confirms this statement:
the left and right panels show the energy localization for 
$\gamma = 50$ and $60$ with the same value of $\epsilon$, respectively.
We clearly see that subdominant peaks merge with the outermost ones.
Therefore, we conclude that the energy localization and the propagation speed of the wall persist even after collisions
if trapping does not occur.

In summary, there are several cases to consider for the energy distribution. 
If the false and true vacua are degenerate, the energy localization is still strong even after the walls propagate over a distance of order bubble radius after collision.
This holds true regardless of whether the scalar field bounces back or not (modified $Z_2$), since the bubble wall does not decelerate after collision.

When the two vacua are not degenerate, the energy distribution depends on the dynamics of the scalar field. 
In case the false vacuum trapping does not occur, the energy localization is still much thinner than the bubble size 
(this has been the case for the hierarchical potential and the quartic potential with $\epsilon = 0.05$).
If the false vacuum trapping occurs after collision,
the scalar field feels a decelerating pressure and the energy gets dispersed from the relativistic front
(as seen for the quartic potential with $\epsilon = 0.5$).
The trapping equation (\ref{eq:trapping}) is hence a useful tool in determining the energy distribution after bubble collisions.

\section{Conclusions}
\label{sec:conc}
\setcounter{equation}{0}

In this paper we studied scalar field bubble collisions in first-order phase transitions in the relativistic regime.
It is of great importance to understand the scalar field dynamics and the energy distribution in this case,
since the shape of the GW spectrum differs significantly depending on whether the bubble walls instantly lose energy at the collision point 
or the energy propagates much further after collision.

We proposed a 'trapping equation' which describes the behavior of bubbles at the initial stage after the collision. 
The equation can be used to determine whether or not the scalar field bounces back and becomes trapped in the false vacuum.
We extensively tested the validity of the trapping equation in a variety of setups in Sec.~\ref{sec:test} and compared with scalar field simulations. 
We also discussed the implication of the trapping equation to $U(1)$ or $SU(2)$ breaking transitions in Sec.~\ref{sec:appl} 
where scalar field simulations are more elaborate.

The false vacuum trapping has a huge impact on the resulting GW spectrum,
since it leads to a decelerating pressure on the propagating scalar field
and therefore changes the extent of energy penetration after collision~\cite{Konstandin:2011ds}.
Ultimately, the 'trapping equation' determines which mechanism of GW production prevails after the phase transition:
The so-called envelope approximation~\cite{Kosowsky:1992rz,Kosowsky:1992vn,Huber:2008hg,Jinno:2016vai} 
or the bulk flow model~\cite{Jinno:2017fby,Konstandin:2017sat}.

\section*{Acknowledgment}

The work of RJ was supported by Grants-in-Aid for JSPS Overseas Research Fellow (No. 201960698).
This work is supported by the Deutsche Forschungsgemeinschaft 
under Germany's Excellence Strategy -- EXC 2121 ,,Quantum Universe`` -- 390833306.

\appendix

\small
\bibliography{ref}

\end{document}